\PassOptionsToPackage{unicode}{hyperref}
\PassOptionsToPackage{hyphens}{url}
\PassOptionsToPackage{dvipsnames,svgnames,x11names}{xcolor}
\documentclass[
  12pt]{article}

\usepackage{amsmath,amssymb}
\usepackage{iftex}
\ifPDFTeX
  \usepackage[T1]{fontenc}
  \usepackage[utf8]{inputenc}
  \usepackage{textcomp} 
\else 
  \usepackage{unicode-math}
  \defaultfontfeatures{Scale=MatchLowercase}
  \defaultfontfeatures[\rmfamily]{Ligatures=TeX,Scale=1}
\fi
\usepackage{lmodern}
\ifPDFTeX\else  
\fi
\IfFileExists{upquote.sty}{\usepackage{upquote}}{}
\IfFileExists{microtype.sty}{
  \usepackage[]{microtype}
  \UseMicrotypeSet[protrusion]{basicmath} 
}{}
\makeatletter
\@ifundefined{KOMAClassName}{
  \IfFileExists{parskip.sty}{%
    \usepackage{parskip}
  }{
    \setlength{\parindent}{0pt}
    \setlength{\parskip}{6pt plus 2pt minus 1pt}}
}{
  \KOMAoptions{parskip=half}}
\makeatother
\usepackage{xcolor}
\makeatletter
\ifx\paragraph\undefined\else
  \let\oldparagraph\paragraph
  \renewcommand{\paragraph}{
    \@ifstar
      \xxxParagraphStar
      \xxxParagraphNoStar
  }
  \newcommand{\xxxParagraphStar}[1]{\oldparagraph*{#1}\mbox{}}
  \newcommand{\xxxParagraphNoStar}[1]{\oldparagraph{#1}\mbox{}}
\fi
\ifx\subparagraph\undefined\else
  \let\oldsubparagraph\subparagraph
  \renewcommand{\subparagraph}{
    \@ifstar
      \xxxSubParagraphStar
      \xxxSubParagraphNoStar
  }
  \newcommand{\xxxSubParagraphStar}[1]{\oldsubparagraph*{#1}\mbox{}}
  \newcommand{\xxxSubParagraphNoStar}[1]{\oldsubparagraph{#1}\mbox{}}
\fi
\makeatother

\newcommand{\citeay}[1]{\citeauthor{#1}, \citeyear{#1}}

\usepackage{longtable,booktabs,array}
\usepackage{calc} 
\usepackage{etoolbox}
\makeatletter
\patchcmd\longtable{\par}{\if@noskipsec\mbox{}\fi\par}{}{}
\makeatother
\IfFileExists{footnotehyper.sty}{\usepackage{footnotehyper}}{\usepackage{footnote}}
\makesavenoteenv{longtable}
\usepackage{graphicx}
\makeatletter
\def\maxwidth{\ifdim\Gin@nat@width>\linewidth\linewidth\else\Gin@nat@width\fi}
\def\maxheight{\ifdim\Gin@nat@height>\textheight\textheight\else\Gin@nat@height\fi}
\makeatother
\setkeys{Gin}{width=\maxwidth,height=\maxheight,keepaspectratio}
\makeatletter
\def\fps@figure{htbp}
\makeatother

\makeatletter
\@ifpackageloaded{caption}{}{\usepackage{caption}}
\AtBeginDocument{%
\ifdefined\contentsname
  \renewcommand*\contentsname{Table of contents}
\else
  \newcommand\contentsname{Table of contents}
\fi
\ifdefined\listfigurename
  \renewcommand*\listfigurename{List of Figures}
\else
  \newcommand\listfigurename{List of Figures}
\fi
\ifdefined\listtablename
  \renewcommand*\listtablename{List of Tables}
\else
  \newcommand\listtablename{List of Tables}
\fi
\ifdefined\figurename
  \renewcommand*\figurename{Figure}
\else
  \newcommand\figurename{Figure}
\fi
\ifdefined\tablename
  \renewcommand*\tablename{Table}
\else
  \newcommand\tablename{Table}
\fi
}
\@ifpackageloaded{float}{}{\usepackage{float}}
\floatstyle{ruled}
\@ifundefined{c@chapter}{\newfloat{codelisting}{h}{lop}}{\newfloat{codelisting}{h}{lop}[chapter]}
\floatname{codelisting}{Listing}

\makeatother
\makeatletter
\@ifpackageloaded{caption}{}{\usepackage{caption}}
\@ifpackageloaded{subcaption}{}{\usepackage{subcaption}}
\makeatother

\ifLuaTeX
  \usepackage{selnolig}  
\fi
\usepackage[]{natbib}
\usepackage{bookmark}
\usepackage{hyperref}

\IfFileExists{xurl.sty}{\usepackage{xurl}}{} 
\hypersetup{
  pdftitle={Title},
  pdfauthor={Author 1; Author 2},
  pdfkeywords={3 to 6 keywords, that do not appear in the title},
  colorlinks=true,
  linkcolor={blue},
  filecolor={Maroon},
  citecolor={Blue},
  urlcolor={Blue},
  pdfcreator={LaTeX via pandoc}}

\newcommand{\anon}{1}

\begin{document}

\def\spacingset#1{\renewcommand{\baselinestretch}%
{#1}\small\normalsize} \spacingset{1}


\if1\anon
{
  \title{\bf Forecasting the Term Structure of Interest Rates with SPDE-Based Models}
  \author{Qihao Duan, Alexandre B. Simas, David Bolin, Rapha\"el Huser \hspace{.2cm}\\
    Statistics Program, CEMSE Division, \\ King Abdullah University of Science and Technology}
  \maketitle
} \fi

\if0\anon
{
  \bigskip
  \bigskip
  \bigskip
  \begin{center}
    {\LARGE\bf Title}
\end{center}
  \medskip
} \fi

\bigskip
\begin{abstract}
The Dynamic Nelson--Siegel (DNS) model is a widely used framework for term structure forecasting. We propose a novel extension that models DNS residuals as a Gaussian random field, capturing dependence across both time and maturity. The residual field is represented via a stochastic partial differential equation (SPDE), enabling flexible covariance structures and scalable Bayesian inference through sparse precision matrices. We consider a range of SPDE specifications, including stationary, non-stationary, anisotropic, and nonseparable models. The SPDE--DNS model is estimated in a Bayesian framework using the integrated nested Laplace approximation (INLA), jointly inferring latent DNS factors and the residual field. Empirical results show that the SPDE-based extensions improve both point and probabilistic forecasts relative to standard benchmarks. When applied in a mean--variance bond portfolio framework, the forecasts generate economically meaningful utility gains, measured as performance fees relative to a Bayesian DNS benchmark under monthly rebalancing. Importantly, incorporating the structured SPDE residual substantially reduces cross-maturity and intertemporal dependence in the remaining measurement error, bringing it closer to white noise. These findings highlight the advantages of combining DNS with SPDE-driven residual modeling for flexible, interpretable, and computationally efficient yield curve forecasting.
\end{abstract}

\noindent%
{\it Keywords:} Dynamic Nelson–Siegel model, latent Gaussian model, Gaussian random field, R-INLA, stochastic differential equation
\vfill

\newpage
\spacingset{1.8} 

\section{Introduction}

The term structure of interest rates, or yield curve, characterizes the relationship between bond yields and maturities at a given point in time and can be formalized as the function $Y(t,m) = -\tfrac{1}{m}\log P(t,m)$, where $Y(t,m)$ denotes the continuously compounded zero-coupon yield at time $t$ for maturity $m$, and $P(t,m)$ is the time-$t$ price of a default-free zero-coupon bond maturing at time $t+m$; equivalently, the instantaneous forward rate is $f(t,m) = -\partial_m \log P(t,m)$ and $Y(t,m) = \tfrac{1}{m}\int_0^{m} f(t,s)\,\rm{d}s$. Thus, $Y(t,m)$ is the average of the forward curve over $[0,m]$ and represents the per-unit-time discount rate that equates a future face value of $1$ at $t+m$ to its present value $P(t,m)$ via the relation $P(t,m) = e^{-Y(t,m)m}$; while this definition relies on continuous compounding, other discrete compounding conventions can be mapped to $Y(t,m)$ through standard algebraic transformations. The yield curve is fundamental in finance, underpinning asset pricing, monetary policy, and risk management because the discount function $P(t,m)$ determines present values across maturities and links the level, slope, and curvature of $Y(t,\cdot)$ to expectations about future short rates \citep{campbell1991yield} and time-varying term premia \citep{fama1987information}. Accurate modeling and forecasting of the yield curve are crucial for financial stability and macroeconomic planning, especially in post-crisis environments marked by unconventional monetary policy and market changes \citep{zhuang2021statistical, congedi2025term, opschoor2025smooth}.

Existing modeling approaches for the yield curve fall broadly into two categories: short-rate models, which describe the evolution of the instantaneous short rate \citep{vasicek1977equilibrium}, and forward-rate models, which specify the dynamics of the entire forward-rate curve. Among the latter, the Nelson--Siegel (NS) model \citep{nelson1987parsimonious} and its dynamic extension by \cite{diebold2006forecasting}, the Dynamic Nelson--Siegel (DNS) model, are particularly influential for their parsimonious representation of yield-curve dynamics through three interpretable factors (level, slope, curvature; see \citeay{diebold2006macroeconomy}). Within the forward-rate framework introduced above, DNS can be viewed as a parametric forward-rate model in which yields are linear combinations of a set of exponentially decaying basis functions in maturity, with the associated factors evolving according to a state-space process; DNS can also be embedded in an arbitrage-free affine setup, but our focus is on its empirical specification.

Numerous extensions of the DNS model have been proposed to enhance its empirical performance. \cite{koopman2010analyzing}, for example, allow the decay parameter and volatility to be time-varying and formulate DNS as a nonlinear state-space system. More recently, \cite{opschoor2025smooth} introduce a smooth shadow-rate version of the DNS model to account for the zero lower bound, delivering improved forecasts under constrained policy regimes. These developments reinforce DNS as a flexible and extensible empirical platform.

Bayesian extensions of the Nelson--Siegel (NS) model have also attracted increasing attention for their ability to incorporate prior information, account for parameter and state uncertainty, and produce full predictive distributions. For example, \cite{carriero2012forecasting} use a Bayesian VAR to improve forecasting of the latent factors; \cite{laurini2014forecasting} adopt a full Bayesian approach to estimate the DNS model with stochastic volatility; and \cite{valente2024bayesian} extend this line further by using a Bayesian long-memory model within the integrated nested Laplace approximation (INLA) framework. These studies illustrate how Bayesian inference enhances the flexibility and forecasting performance of NS-type models.

INLA (\cite{rue2009approximate}) is widely used across a variety of statistical domains, including DNS-related applications, as a method for approximate Bayesian inference in latent Gaussian models. Unlike Markov Chain Monte Carlo methods, INLA provides fast, accurate, deterministic approximations to posterior marginals by exploiting sparsity of the precision structure of the latent Gaussian components. INLA has been increasingly adopted across applied fields, including yield-curve modeling. For example, \cite{valente2024bayesian} estimate a long-memory DNS model using INLA with a weighted linear predictor, demonstrating that INLA can be successfully applied to this class of dynamic term-structure models.

Although these INLA-based DNS model typically delivers strong forecasting performance, its gains are often modest because it does not account for residual correlations across time and maturity. Several recent papers introduce geostatistical methods for modeling the term structure in two dimensions (see \citeay{arbia2015forecasting}). \cite{congedi2025term} adopt a kriging-based approach and model the residual surface as a bivariate random field with a nonseparable covariance structure.

While such methods offer a valuable direction, they remain rooted in covariance-based inference, which can be computationally intensive and difficult to scale, owing to dense covariance matrices and costly matrix factorizations, especially for large datasets or nonseparable covariance structures. By contrast, we propose a stochastic partial differential equation (SPDE)-based residual model, enabling fast and flexible Bayesian inference via INLA. This approach not only improves computational efficiency but also facilitates richer model extensions, including anisotropy, non-stationarity, and nonseparable interactions.

In this paper, we propose a new approach for forecasting the term structure of interest rates by extending the Dynamic Nelson–Siegel (DNS) model to account for residual structure that is not captured by the standard latent factors. Specifically, we assume that the residuals contain systematic dependence across time and maturity, which we model as a Gaussian random field with a flexible covariance structure. Our aim is to represent this structure in a way that balances both modeling flexibility and computational tractability. To that end, we draw on the success of the Matérn covariance family, which is widely used in spatial statistics and machine learning for its tunable smoothness and locality properties \citep{seeger2004gaussian, van2024low}, and, as emphasized by \cite{stein1999interpolation}, is practically important because its smoothness parameter can be estimated from data. 

We model residual dependence in the yield curve using SPDE-based Gaussian random field models. Our best specification is a nonseparable spatio-temporal SPDE model, which distinguishes calendar time from maturity and captures both serial dependence and cross-sectional correlation. For comparison, we also consider several ``spatial`` SPDE alternatives, in which a Gaussian random field is defined on the joint time–maturity domain treated as a two-dimensional spatial domain, including stationary, non-stationary, and anisotropic specifications. For the ``spatial`` SPDE specifications, we implement the rational SPDE framework of \cite{bolin2024covariance}, which allows the smoothness parameter of the associated Matérn covariance structure to take non-integer values and to be estimated directly from the data. We evaluate the empirical performance of these specifications in yield curve forecasting and their economic value in a mean–variance bond-portfolio setting, where better forecasts translate directly into higher risk-adjusted investment performance.

Another common limitation of all the DNS framework is that the decay parameter $\lambda$ is typically fixed at a predetermined value or estimated in a preliminary step, rather than jointly with the latent factors, which precludes full uncertainty quantification for $\lambda$ and can bias the factor loading estimates. We address this limitation by treating $\lambda$ as part of the latent model structure and estimating it jointly with the dynamic factors and the SPDE residual field. This is facilitated by the \texttt{inlabru} framework \citep{lindgren2024inlabru}, which extends INLA to accommodate nonlinear predictors through iterative {linearization}. We also {consider} alternative prior specifications for $\lambda$ and {assess} their impact on forecasting performance. This joint estimation also enables a coherent evaluation of the model’s economic value, allowing us to quantify how improved factor loadings and residual modeling enhance portfolio decisions in practice. In summary, in contrast to existing INLA-based DNS models, we introduce two key extensions. We model the DNS residuals using an SPDE-based Gaussian random field to capture structured dependence across time and maturity, and we jointly estimate the Nelson–Siegel decay parameter together with the latent factors, allowing its uncertainty to be fully propagated.

An important finding from our analysis is that modeling residual dependence is also crucial for inference on the Nelson–Siegel factors themselves. When residual correlations across time and maturity are ignored, the estimated latent factors can absorb systematic variation that is not truly factor-driven. Allowing for a structured residual component can therefore lead to substantially different estimates of the level, slope, and curvature dynamics. This distinction is economically relevant, as it affects how yield-curve movements are interpreted and how factor-based forecasts are used in practice.

The paper proceeds as follows. Section \ref{sec:methodology} describes the proposed methodological framework including the various SPDE-DNS models and the Bayesian inference approach based on INLA. Section \ref{sec:numerical} describes the results of our data analysis and outlines the economic-value evaluation design, including the mean--variance bond-portfolio framework and the performance-fee metric used to quantify utility gains relative to a Bayesian DNS benchmark. Section \ref{sec:conclusion} concludes with a discussion and outlines further research directions. The R code that implements the SPDE-DNS models and reproduces the analysis in this paper is available at the following GitHub repository: \url{https://github.com/qlhlduan/SPDE-DNS}.

\section{Methodology}\label{sec:methodology}

In this section, we first introduce the classical Nelson--Siegel model for the yield curve, and its dynamic extension. Employing Bayesian inference via the R-INLA framework to quantify parameter uncertainty, we formulate it as a latent Gaussian model with a Gaussian observation equation: the latent field comprises the DNS factors together with the SPDE residual effect, and yields are observed with idiosyncratic Gaussian noise. We then further discuss the proposed residual-adjusted extensions based on the rational SPDE approach.

\subsection{Dynamic Nelson–Siegel (DNS) model}
The DNS model, introduced by \cite{diebold2006forecasting}, extends the original Nelson--Siegel yield curve representation by allowing its parameters to evolve over time as latent stochastic processes. Let $t\in\{1,\ldots,T\}$ index observation times and let $m\in\{m_1,\ldots,m_M\}$ denote the set of maturities (e.g., in months) observed at each $t$.

The DNS observation equation specifies the continuously compounded zero-coupon yield $Y(t,m)$ at time $t$ and maturity $m$ as
\begin{align}\label{dns}
    &Y(t,m) \;=\; Z(t,m) \;+\; \epsilon_{t,m}, \qquad t=1,\ldots,T,\;\; m\in\{m_1,\ldots,m_M\},
\end{align}
where $Z(t,m) \;=\; \beta_{1,t} + \beta_{2,t}\left(\frac{1-e^{-\lambda_t m}}{\lambda_t m}\right) + \beta_{3,t}\left(\frac{1-e^{-\lambda_t m}}{\lambda_t m} - e^{-\lambda_t m}\right)$, and $\epsilon_{t,m}$ is an idiosyncratic measurement error assumed to be independent and identically distributed (i.i.d.) across $t$ and $m$ and independent of all latent processes. The three time-varying factors $(\beta_{1,t},\beta_{2,t},\beta_{3,t})$ in \eqref{dns} are latent Gaussian processes over time that control the level, slope, and curvature of the yield curve, respectively, while $\lambda_t$ determines the exponential decay of these factors.

The loading on the level factor is identically one, so $\beta_{1,t}$ shifts the entire curve up and down. The slope loading starts near one at short maturities and decays monotonically to zero as $m$ increases, so $\beta_{2,t}$ captures the short-vs-long steepness of the yield curve, reflecting how quickly short-term rates revert toward longer-term levels. The curvature loading starts at zero at short maturities, peaks at intermediate maturities, and returns toward zero at long maturities, allowing $\beta_{3,t}$ to generate hump-shaped deviations.

The factor dynamics are modeled as independent AR(1) processes with drifts, i.e.,
\[
\begin{pmatrix}
\beta_{1,t}-\mu_1\\[2pt]
\beta_{2,t}-\mu_2\\[2pt]
\beta_{3,t}-\mu_3
\end{pmatrix}
=
\begin{pmatrix}
\phi_1 & 0 & 0\\
0 & \phi_2 & 0\\
0 & 0 & \phi_3
\end{pmatrix}
\begin{pmatrix}
\beta_{1,t-1}-\mu_1\\[2pt]
\beta_{2,t-1}-\mu_2\\[2pt]
\beta_{3,t-1}-\mu_3
\end{pmatrix}
+
\begin{pmatrix}
\delta_{1,t}\\[2pt]
\delta_{2,t}\\[2pt]
\delta_{3,t}
\end{pmatrix},
\qquad
\delta_{i,t}\stackrel{\text{i.i.d.}}{\sim}\mathcal{N}(0,\sigma_{\beta_i}^2),
\]
with $|\phi_i|<1$ for stationarity, $(\mu_1, \mu_2,\mu_3)$ are the unconditional means of the level, slope, and curvature factors, and $\{\delta_{i,t}\}$ is normally distributed noise, with $\sigma_{\beta_i}$ controlling the volatility of innovations and thus the month-to-month variability in each factor. Thus, $(\beta_{1,t},\beta_{2,t},\beta_{3,t})$ form a latent Gaussian state vector evolving over time.

The decay parameter $\lambda_t$ governs the rate at which the slope and curvature loadings decay with maturity and thereby controls both the location of the curvature peak and the speed at which the slope effect diminishes. In most empirical implementations, including \cite{diebold2006forecasting}, $\lambda_t$ is treated as constant over time, i.e., $\lambda_t\equiv\lambda$. A common choice is $\lambda=0.0609$, which places the curvature peak at a maturity of around 30 months. We adopt this convention when we refer to a ``fixed-$\lambda$'' DNS specification and later consider joint estimation of $\lambda$ as part of the latent structure. They estimate and forecast the yield curve using a two-step procedure: first, for each $t$, they regress $Y(t,\cdot)$ on the three maturity loadings (given the value of $\lambda$) to obtain estimates of $(\beta_{1,t},\beta_{2,t},\beta_{3,t})$; second, they fit separate AR(1) models to the factor time series and use them to produce factor forecasts; finally, they map those factor forecasts through the loadings to obtain $\widehat{Y}(t+1,m)$.

As our first benchmark model, we embed this framework with a fixed decay parameter (e.g., $\lambda=0.0609$) in a latent Gaussian setting. Treating $\lambda$ as fixed renders the loading functions deterministic, allowing us to express the model in vector form as $Y(t,m) = \boldsymbol{A} \boldsymbol{\beta}_t + \epsilon_{t,m}$, where $\boldsymbol{\beta}_t = (\beta_{1,t}, \beta_{2,t}, \beta_{3,t})^\top$ represents the latent state vector of level, slope, and curvature factors, $\boldsymbol{A}$ is the so called observation matrix and $\epsilon_{t,m}$ is unstructured Gaussian noise.

Within this structure, the observation matrix $\boldsymbol{A}$ plays the critical role of mapping the latent factors to the observed yields. It is a fixed, deterministic matrix whose rows correspond to the observed maturities $m_1, \dots, m_M$, and whose columns contain the corresponding Nelson--Siegel loadings for the three factors: i.e.,
\[
\boldsymbol{A} = 
\begin{bmatrix}
1 & \frac{1-e^{-\lambda m_1}}{\lambda m_1} & \left(\frac{1-e^{-\lambda m_1}}{\lambda m_1} - e^{-\lambda m_1}\right) \\
1 & \frac{1-e^{-\lambda m_2}}{\lambda m_2} & \left(\frac{1-e^{-\lambda m_2}}{\lambda m_2} - e^{-\lambda m_2}\right) \\
\vdots & \vdots & \vdots \\
1 & \frac{1-e^{-\lambda m_M}}{\lambda m_M} & \left(\frac{1-e^{-\lambda m_M}}{\lambda m_M} - e^{-\lambda m_M}\right)
\end{bmatrix}.
\]
In the INLA framework, $\boldsymbol{A}$ precisely functions as the projection matrix that defines the linear combination of the latent states ($\boldsymbol{\beta}_t$) for each data point ($Y(t,m)$). With $\lambda$ fixed, $\boldsymbol{A}$ depends solely on the maturities, creating a standard linear predictor $\boldsymbol{\eta}_t = \boldsymbol{A}\boldsymbol{\beta}_t$ that fits directly into the R-INLA machinery, which will be introduced in Section~\ref{bayes}. This formulation provides a properly specified model, that we denote as DNS–AR(1), and which serves as the baseline model before extending the framework to jointly estimate $\lambda$ and the SPDE-based residual field in subsequent sections.

\subsection{Residual-adjusted SPDE models}\label{models}
\subsubsection{General model specification}
In this section, we introduce a more flexible specification for the residual term, which is often assumed to be white noise. We generalize the residual term to allow structured dependence and rewrite the yield function $Y(t,m)$ on a bounded domain $\mathcal{D}\subset\mathbb{R}^2$ spanned by calendar time $t$ and maturity $m$ as
\begin{align}
    &Y(t,m) \;=\; Z(t,m) \;+\; \varepsilon_{t,m}, \qquad t = 1,\ldots,T,\;\; m = 1,\ldots,M,
    \label{obsmodel}
\end{align}
where $ \varepsilon_{t,m} = u(t,m) + e_{t,m}, e_{t,m}\stackrel{\text{i.i.d.}}{\sim}\mathcal{N}(0,\sigma^2), e_{t,m} \perp\!\!\!\perp u(t,m)$, so that $\boldsymbol{\varepsilon}=(\varepsilon_{t,m})_{t=1,m=1}^{T,M}\sim \mathcal{N}\!\big(\boldsymbol{0},\;\boldsymbol{\Sigma} + \sigma^2 \boldsymbol{I}\big)$. In this model, $Z(t,m)$ is the DNS trend component and $\boldsymbol{\Sigma}$ is the covariance matrix induced by the zero-mean Gaussian random field $u(t,m)$ over $\mathcal{D}$, capturing dependence across both calendar time and maturity (i.e., covariances between different times and across different maturities). In other words, we decompose the residual term into a structured component $u(t,m)$ and an idiosyncratic white-noise term $e_{t,m}$.

To model $u(t,m)$, we employ a class of SPDE-based residual processes, introduced in the following section, constructed via the rational SPDE approach of \cite{bolin2020rational}. This framework provides a flexible family of covariance structures, with tunable smoothness, anisotropy, and nonstationarity, while retaining computational scalability by working with sparse precision matrices arising from SPDE discretizations.

\subsubsection{``Spatial'' models}
In classical spatial statistics, the Matérn covariance structure, defined as
\[
C(h) \;=\; \frac{\sigma^2}{2^{\nu-1}\Gamma(\nu)}\,(\kappa\,h)^\nu\,K_\nu(\kappa\,h), \qquad h > 0,
\]
with $\Gamma(\cdot)$ the gamma function, and $K_\nu(\cdot)$ the modified Bessel function of the second kind, is widely for Gaussian-process modeling. Its main appeal is that it because it flexibly captures correlation decay with distance $h$ thanks to the range parameter $\kappa > 0$ and it admits a tunable smoothness parameter $\nu$ that controls path regularity from very rough to arbitrarily smooth (with the squared-exponential limit approached as $\nu\to\infty$). Another key aspect of the Mat\'ern covariance model is that it has a natural link with SPDEs, as explained below, which facilitates complex spatial modeling in high dimensions.

We first consider a class of residual models that treat the joint time--maturity domain $D\subset\mathbb{R}^2$ as a ``spatial'' domain, i.e., without distinguishing between the specific roles of the temporal and maturity axes. In this setting, the structured residual component $u(t,m)$ is represented as the weak solution to the SPDE:
\begin{equation}
    L(\cdot)^{\alpha/2} (\tau u(\cdot)) \;=\; W(\cdot), \quad \hbox{on } D,
    \label{operator}
\end{equation}
where $L$ is a second order differential operator, $\tau>0$ scales the marginal variance and $\alpha>0$ relates to the Matérn smoothness via $\alpha=\nu + d/2$ with spatial dimension $d=2$. The driving term $W(\cdot)$ denotes Gaussian white noise on $D$ (a generalized random field); see Appendix~A for a brief review of Gaussian white noise and weak solutions to such SPDEs. Different choices of the operator structure in $L$ yield the stationary, nonstationary, and anisotropic models discussed below. 

Direct covariance-based simulation and inference with Matérn fields can be computationally prohibitive due to dense covariance matrices. The SPDE approach circumvents this by discretizing the domain (e.g., via the finite element method, see Appendix~B), which yields sparse precision matrices and enables scalable inference with sparse linear-algebra routines (e.g., sparse Cholesky factorizations in INLA).

\textbf{Stationary isotropic model} 
A stationary isotropic Gaussian random field $u(t,m)$ on $D$ corresponds to the operator $L = (\kappa - \Delta)^{\alpha/2}$ in \eqref{operator}, where $\kappa>0$ controls the correlation range and $\alpha=\nu+d/2$ determines smoothness via $\nu>0$. On an infinite domain, the stationary solution to \eqref{operator} corresponds to a Gaussian random field with Matérn covariance function $C(h)$, where $h = \| (t_1,m_1)-(t_2,m_2)\|$ and $\| \cdot \|$ denotes the Euclidean norm on $D$. In the present setting, the field is defined on a bounded time--maturity domain, and the resulting covariance is nearly indistinguishable from the stationary Matérn covariance away from the domain boundaries. 

\textbf{Nonstationary model}
While the stationary model provides a natural baseline, constant range and variance can be too restrictive for yield-curve residuals. To allow for heterogeneity, we consider a nonstationary SPDE model in which both the local correlation range $\rho(t,m)$ and marginal variance $\sigma(t,m)$ vary smoothly over the domain, as detailed in \cite{lindgren2015bayesian}. These parameters are are directly related to the SPDE parameters, where \begin{align*}
    \log \tau = \frac{1}{2} \log(\frac{\Gamma(\nu)}{\Gamma(\alpha)(4\pi})^{d/2}) - \log \sigma - \nu \log \kappa, \quad \log \kappa = \frac{\log(8\nu)}{2} - \log \rho.
\end{align*}  To ensure positivity and numerical stability, $\rho(t,m)$ and  $\sigma(t,m)$ are modeled on the logarithmic scale and allowed to vary smoothly over time $t$ and maturity $m$. The non-stationary parameterization is specified as
\begin{align*}
\log \rho(t,m) = \gamma_0 + \gamma_2 t + \gamma_4 m, \quad
\log \sigma(t,m) = \gamma_1 + \gamma_3 t + \gamma_5 m,
\end{align*}
with regression coefficients $\gamma_i$, $i=0,\ldots,5$ to estimate. This specification permits the correlation range and marginal variance to vary smoothly across the time–maturity domain.

\textbf{Anisotropic model}
In the yield-curve context, isotropy is often unrealistic: correlations across calendar time reflect macroeconomic dynamics, while dependencies across maturity reflect cross-sectional pricing effects. This anisotropic SPDE formulation is based on the deformation approach introduced by \cite{fuglstad2015exploring}. To capture directional asymmetries, we replace the Laplacian with a general second-order elliptic operator defined by a positive-definite $2\times2$ matrix $\boldsymbol{H}$. That is, we use $L = (\kappa - \nabla\!\cdot \boldsymbol{H} \nabla)^{\alpha/2} \ \hbox{in} \ \eqref{operator}$.

This deformation induces elliptical (rather than circular) correlation contours, allowing different effective ranges along the time and maturity directions. Equivalently, the induced Matérn covariance depends on the anisotropic distance $\sqrt{\boldsymbol{h}^\top \boldsymbol{H}^{-1}\boldsymbol{h}}$, where $\boldsymbol{h}$ is the separation vector in $D$. While this improves flexibility, identifiability considerations (e.g., between $\boldsymbol{H}$ and the marginal variance) call for careful regularization, through constrained parameterization of $\boldsymbol{H}$ and informative priors for SPDE parameters.

\textbf{Rational approximation}
A limitation of the basic SPDE construction is that it traditionally requires $\alpha\in\mathbb{N}$ in \eqref{operator}, restricting smoothness to discrete values. This can be addressed using the rational SPDE approach, which introduces a rational approximation to the covariance function of the solution to the SPDEs \eqref{operator}, thereby enabling inference for arbitrary smoothness while retaining computational tractability (see Appendix~C). All three spatial models are implemented via this rational approximation framework, which permits estimation of smoothness which is not necessarily an integer.

In summary, our proposed stationary, nonstationary, and anisotropic specifications form a unified class of fractional spatial SPDE models on $D\subset\mathbb{R}^2$, differing primarily in how $\kappa(\cdot)$, $\tau(\cdot)$, and the differential operator are parameterized. While these models flexibly capture residual dependence, they remain fundamentally ``spatial'' in nature because they model time and maturity jointly through a single spatial operator. In contrast, the nonseparable ``spatio-temporal'' SPDE model introduced next treats maturity as the spatial domain and time as an explicit dynamic dimension, thereby capturing both cross-sectional dependence across maturities and serial dependence over time within a unified operator framework.

\subsubsection{Nonseparable ``spatio-temporal'' model}

In contrast to the purely spatial specifications above, the spatio-temporal SPDE model treats maturity $m \in M$ as the spatial domain and calendar time $t \in T$ as the temporal dimension. This formulation allows the residual process to evolve dynamically over time while preserving structured dependence across maturities.

To incorporate temporal dynamics directly into the SPDE operator, we consider a spatio-temporal extension of the Matérn-type spatial model, in which the temporal evolution is governed by a first-order differential operator and spatial dependence is controlled through fractional powers of the Laplacian (\citeay{lindgren2020diffusion}). Specifically, we adopt the following spatio-temporal SPDE:
\begin{equation}
    \left(\frac{\partial}{\partial t} + \gamma (\kappa^2 - \Delta_m)^{\alpha} \right)u(\cdot) = \mathrm{d}W_Q(\cdot),
    \quad \text{on} \quad T \times M.
    \label{stem}
\end{equation}

Here, $\kappa > 0$ is the spatial range parameter and $\gamma > 0$ governs the temporal rate of evolution of the spatial field. The driving noise $W_Q$ is a $Q$-Wiener process (see Appendix~D)with spatial covariance operator
$\sigma^2 (\kappa^2 - \Delta_m)^{-\beta}$, where $\sigma^2$ denotes the marginal variance. The parameters $\alpha$ and $\beta$ control the smoothness of the spatial operator and the noise process, respectively.

This specification yields a nonseparable spatio-temporal residual process $u(t,m)$ that evolves continuously over time while maintaining flexible spatial dependence across maturities. While the temporal smoothness is fixed by the first-order time derivative, the model provides sufficient flexibility to capture persistent cross-maturity dependence and dynamic residual structure in yield curve data.

\subsection{Bayesian Inference}\label{bayes}
The \texttt{R-INLA} package provides accurate and computationally efficient Bayesian inference for latent Gaussian models by exploiting the sparsity of precision matrices, leading to substantial computational gains over simulation-based methods such as Markov chain Monte Carlo. Since all models considered in this paper can be formulated as latent Gaussian models, INLA offers a natural and scalable inference framework. Such models are defined hierarchically through three components: a set of hyperparameters $\boldsymbol{\theta}$, a latent Gaussian field $\boldsymbol{x}$, and the likelihood function of the model for the observations $\boldsymbol{y} =(y_i)$ given $\boldsymbol{x}$. This model can be written as: 
\begin{align*}
    \boldsymbol{\theta} &\sim \pi(\boldsymbol{\theta})    & \text{hyperparameters},   \\
   \boldsymbol{x} \mid \boldsymbol{\theta} &\sim N(\boldsymbol{0}, \boldsymbol{Q}(\boldsymbol{\theta})^{-1}) \ &\text{latent field}, \\
   \boldsymbol{y} \mid \boldsymbol{x}, \boldsymbol{\theta} &\sim \prod_{i} \pi(y_i \mid \eta_i(\boldsymbol{x}), \boldsymbol{\theta}) &\text{likelihood},
\end{align*}
where $\pi(\boldsymbol{\theta})$ is the prior distribution for the hyperparameters, and $\boldsymbol{Q}(\boldsymbol{\theta})$ is the precision (inverse covariance) matrix of the latent field $\boldsymbol{x}$. The linear predictor $\boldsymbol{\eta} =(\eta_i(\boldsymbol{x}))$ connects the latent field to the observations. This link is defined by the linear projection $\boldsymbol{\eta}(\boldsymbol{x}) = \boldsymbol{A}\boldsymbol{x}$, where the observation matrix $\boldsymbol{A}$ maps the latent vector $\boldsymbol{x}$ to the predictor $\eta_i$ for each observation $y_i$, found as the $i$-th component of $\boldsymbol{\eta}(\boldsymbol{x})$.

For any arbitrary point $\boldsymbol{x}^*$ (typically chosen as the posterior mean $\boldsymbol{\mu}_{\boldsymbol{x}|\boldsymbol{y}}$), we can write:
\[
\pi(\boldsymbol{y} \mid \boldsymbol{\theta}) = \frac{\pi(\boldsymbol{x} \mid \boldsymbol{\theta})\pi(\boldsymbol{y} \mid \boldsymbol{x}, \boldsymbol{\theta})}{\pi(\boldsymbol{x} \mid \boldsymbol{y}, \boldsymbol{\theta})}\bigg|_{\boldsymbol{x} = \boldsymbol{x}^*}.
\]
In the special case where the observation model is Gaussian, as in our yield curve application, the likelihood $\pi(\boldsymbol{y} \mid \boldsymbol{x}, \boldsymbol{\theta})$ takes a particularly simple form. In this case, the nested Laplace approximation that INLA employs for general non-Gaussian likelihoods is not needed, and the log marginal likelihood can be evaluated directly by substituting the log-densities:
\begin{align*}
\log \pi(\boldsymbol{y} \mid \boldsymbol{\theta}) &= -\frac{n}{2} \log(2\pi) 
+ \frac{1}{2} \log \det(\boldsymbol{Q}(\boldsymbol{\theta}))
+ \frac{1}{2} \log \det(\boldsymbol{Q}_\varepsilon)
- \frac{1}{2} \log \det(\boldsymbol{Q}_{\boldsymbol{x}|\boldsymbol{y}}) \\
&\quad - \frac{1}{2} (\boldsymbol{\mu}_{\boldsymbol{x}|\boldsymbol{y}}^\top \boldsymbol{Q}(\boldsymbol{\theta}) \boldsymbol{\mu}_{\boldsymbol{x}|\boldsymbol{y}})
- \frac{1}{2} (\boldsymbol{y} - \boldsymbol{A}_{\boldsymbol{x}} \boldsymbol{\mu}_{\boldsymbol{x}|\boldsymbol{y}})^\top \boldsymbol{Q}_\varepsilon (\boldsymbol{y} - \boldsymbol{A}_{\boldsymbol{x}} \boldsymbol{\mu}_{\boldsymbol{x}|\boldsymbol{y}}).
\end{align*}
where the posterior distribution of $\boldsymbol{x}$ given $\boldsymbol{y}$ is $\boldsymbol{x} \mid \boldsymbol{y} \sim \mathcal{N}(\boldsymbol{\mu}_{\boldsymbol{x}|\boldsymbol{y}}, \boldsymbol{Q}_{\boldsymbol{x}|\boldsymbol{y}}^{-1}),$
where $\boldsymbol{Q}_{\boldsymbol{x}|\boldsymbol{y}} = \boldsymbol{A}^\top \boldsymbol{Q}_\varepsilon \boldsymbol{A} + \boldsymbol{Q}, \text{and} \  \boldsymbol{\mu}_{\boldsymbol{x}|\boldsymbol{y}} = \boldsymbol{Q}_{\boldsymbol{x}|\boldsymbol{y}}^{-1} \boldsymbol{A}^\top \boldsymbol{Q}_\varepsilon \boldsymbol{y}.$

Using the conditional marginal likelihood $\pi(y_i \mid \boldsymbol{\theta})$, the posterior distribution of the hyperparameters $\boldsymbol{\theta}$ given the data $\boldsymbol{y}$ is expressed as:
\[
\pi(\boldsymbol{\theta} \mid \boldsymbol{y}) = \frac{\pi(\boldsymbol{\theta}, \boldsymbol{y})}{\pi(\boldsymbol{y})} = \frac{\pi(\boldsymbol{\theta})\pi(\boldsymbol{y} \mid \boldsymbol{\theta})}{\pi(\boldsymbol{y})}.
\]

Recall that the yield rate is modeled as
\begin{align} \label{fullmodel}
Y(t,m) = Z(t,m) + \varepsilon_{t,m} = Z(t,m) + u(t,m) + e_{t,m},
\end{align}

where $\boldsymbol{e}=(e_{t,m})_{t=1,m=1}^{T,M} \sim \mathcal{N}(\boldsymbol{0}, \boldsymbol{Q}_e^{-1}), (t,m) \in \mathcal{D},$ with $\boldsymbol{Q}_e$ a sparse precision matrix. Note that it is reasonable to assume that $\boldsymbol{Q}_e = \sigma_e^{-2} \boldsymbol{I}$ is a diagonal matrix after the correlation structure has been captured by $u(t,m)$. 

The residual-adjusted model \eqref{fullmodel} can be expressed as: $\boldsymbol{y} = \boldsymbol{A}_x \boldsymbol{x}+ \boldsymbol{A}_u \boldsymbol{u} + \boldsymbol{e},$
where $\boldsymbol{A}_x$ and $\boldsymbol{A}_u$ are projection matrices linking the latent components to the observations. Specifically, $\boldsymbol{A}_x$ maps the latent level, slope, and curvature factors to the observed yields through the corresponding maturity loadings, and $\boldsymbol{A}_u$ links the SPDE latent field to the observations by evaluating the residual process at each location. To fit this model within the INLA framework, we introduce the augmented latent vector $\tilde{\boldsymbol{x}} = (\boldsymbol{x}, \boldsymbol{u})^{\top}$ and the block diagonal observation matrix $\boldsymbol{A} = \text{diag}(\boldsymbol{A}_x, \boldsymbol{A}_u)$. This reformulation yields a model of the form $\boldsymbol{y} = \boldsymbol{A}\tilde{\boldsymbol{x}} + \boldsymbol{e}$, which directly fits the standard INLA structure. Here, $\boldsymbol{x}$ stems from a latent AR(1) process, $\boldsymbol{u}$ is the SPDE term, and $\boldsymbol{e}$ is the i.i.d. Gaussian observation noise.

For the latent AR(1) term $\boldsymbol{x} = (x_1,\ldots,x_n)^\top$, it is natural to formulate its dynamics as a Gaussian Markov random field. The AR(1) process is here defined as
\begin{align*}
    x_1 &\sim \mathcal{N}\Bigl(0,\frac{1}{\tau(1-\phi^2)}\Bigr), \\
x_t \mid x_{t-1} &\sim \mathcal{N}\Bigl(\phi\, x_{t-1}, \frac{1}{\tau}\Bigr), \quad t=2,\dots,n,
\end{align*}
where $\phi$ is the autoregressive coefficient with $|\phi|<1$, and $\tau$ is the precision parameter. For computational reasons, INLA uses the following parameterization: $\tau = \exp(\theta_1), \phi = 1/(1+ \exp{(-\theta_2)})$. Therefore, $\boldsymbol{Q}(\boldsymbol{\theta})$, the precision matrix of the latent field $\boldsymbol{x} | \boldsymbol{\theta}$ is a sparse block-diagonal precision matrix $ \boldsymbol{Q}(\boldsymbol{\theta}) = \operatorname{diag}(\boldsymbol{Q}_Y, \boldsymbol{Q}_u),$
where $\boldsymbol{Q}_Y$ is a sparse precision matrix of AR(1) dynamics and $\boldsymbol{Q}_u$ is the sparse precision matrix of the SPDE terms, for example the stationary Gaussian random field. 

In addition, we employ the \texttt{rSPDE} package to construct and manipulate fractional SPDE models within the INLA framework. The package provides efficient finite element representations of fractional differential operators, including rational approximations of non-integer powers, enabling flexible control over the smoothness of the latent Gaussian field. Used in conjunction with \texttt{R-INLA}, \texttt{rSPDE} allows scalable Bayesian inference for Matérn-type models with arbitrary smoothness parameters.

\subsection{Latent Non-linear Predictors with \texttt{inlabru}}
Despite its advantages, a central limitation of standard INLA is that the observation matrix $\boldsymbol{A}$ must be fixed and cannot depend on unknown parameters. In the case where one wants to estimate the decay parameter $\lambda$ from data, however, the joint posterior has an observation matrix $\boldsymbol{A}$ that depends on the unknown $\lambda$ parameter, as it is of the form:
\[
\pi(\boldsymbol{\beta}, \boldsymbol{u}, \boldsymbol{\theta}, \lambda \mid \boldsymbol{y}) \propto 
   \pi(\boldsymbol{\theta})\,\pi(\boldsymbol{\beta},\boldsymbol{u} \mid \boldsymbol{\theta})\,\pi(\boldsymbol{y} \mid \boldsymbol{A}(\lambda)\boldsymbol{\beta} + \boldsymbol{u}, \boldsymbol{\theta}).
\]
Because of this, we now propose an extension to the standard INLA estimation method that allows fully Bayesian inference for all model components and hyperparameters, including $\lambda$.

The \texttt{inlabru} package \citep{lindgren2024inlabru} extends the \texttt{INLA} framework to models with non-linear predictors by introducing an iterative linearisation scheme. Specifically, under the same notation as the previous subsection, for a non-linear predictor $\boldsymbol{\eta}(\boldsymbol{x})$, a first-order Taylor expansion around a linearisation point $\boldsymbol{x}_{0}$ yields an expression of the form
\[
\boldsymbol{\eta}(\boldsymbol{x}) 
= \boldsymbol{\eta}(\boldsymbol{x}_{0}) 
+ \sum_{j} \boldsymbol{B}^{(j)}\bigl(\boldsymbol{x}^{(j)} - \boldsymbol{x}_{0}^{(j)}\bigr) 
= \left[\boldsymbol{\eta}(\boldsymbol{x}_{0}) - \sum_{j}\boldsymbol{B}^{(j)}\boldsymbol{x}_{0}^{(j)}\right] 
+ \sum_{j}\boldsymbol{B}^{(j)}\boldsymbol{x}^{(j)} 
= \boldsymbol{\delta} + \sum_{j}\boldsymbol{B}^{(j)}\boldsymbol{x}^{(j)}.
\]
In this formulation, $\boldsymbol{x}$ denotes the full vector of latent Gaussian variables entering the predictor, which may consist of multiple components $\boldsymbol{x}^{(j)}$, such as AR(1) dynamic factors or latent random fields. The linearisation point $\boldsymbol{x}_{0}$ (with $\boldsymbol{x}_0^{(j)}$ for each component) represents the current estimate of these latent variables, typically obtained from the previous iteration of the algorithm. The matrices $\boldsymbol{B}^{(j)}$ collect the partial derivatives of the non-linear predictor $\boldsymbol{\eta}(\boldsymbol{x})$ with respect to the corresponding latent components $\boldsymbol{x}^{(j)}$, evaluated at $\boldsymbol{x}_{0}$. The resulting approximation expresses the non-linear predictor as a linear function of $\boldsymbol{x}$ plus an offset term $\boldsymbol{\delta}$, which ensures that the approximation is locally exact at the linearisation point and therefore must be updated at each iteration.

However, despite this flexibility, \texttt{inlabru} still inherits the key restriction of \texttt{INLA}: parameters that appear inside the observation matrix $\boldsymbol{A}$ (such as the decay parameter $\lambda$ in our model) cannot be directly estimated, since $\boldsymbol{A}$ must be fixed. To overcome this limitation, we propose a novel approach that does not treat $\lambda$ as a fixed parameter in the observation matrix, but instead embeds it within the non-linear predictor itself, enabling joint estimation with other latent variables. Specifically, we consider the desired predictor $\boldsymbol{\eta}(\boldsymbol{x}) = \boldsymbol{A}(\lambda)\,\boldsymbol{\beta} + \boldsymbol{u},$ which contains two layers of nonlinearity:
first, the loadings matrix $\boldsymbol{A}(\lambda)$ depends on the decay parameter $\lambda$ through the Nelson–Siegel functions $\boldsymbol{\beta}(\lambda) = (1, \ \frac{1 - e^{-\lambda m}}{\lambda m}, \ \frac{1 - e^{-\lambda m}}{\lambda m} - e^{-\lambda m})^\top.$ Second, there is a multiplicative interaction between the loadings and the coefficients, so that $h(\boldsymbol{A}(\lambda), \boldsymbol{\beta}) = \boldsymbol{A}(\lambda)\boldsymbol{\beta}$. Instead of treating $\lambda$ as a fixed parameter within $\boldsymbol{A}(\lambda)$, we reparametrize it as a Gaussian latent variable $\tilde{\lambda}$ that enters the nonlinear predictor via the transformation
\[
\lambda = g_{\boldsymbol{\theta}}\bigl(\Phi(\tilde{\lambda})\bigr),
\]
where $\Phi$ is the standard normal cumulative distribution function, and $g_{\boldsymbol{\theta}}$ denotes the quantile function of a chosen prior distribution for $\lambda$. This transformation allows us to specify flexible priors for $\lambda$ while keeping $\tilde{\lambda}$ Gaussian, making it directly compatible with the latent Gaussian structure of \texttt{inlabru}.

Under this formulation, the predictor becomes
\[
\boldsymbol{\eta}(\boldsymbol{x}) = \boldsymbol{A}\bigl(g_{\boldsymbol{\theta}}(\Phi(\tilde{\lambda}))\bigr)\,\boldsymbol{\beta} + \boldsymbol{u} = \boldsymbol{A}(\lambda)\boldsymbol{\beta} + \boldsymbol{u},
\]
and is handled by \texttt{inlabru} through iterative linearisation with respect to all latent variables, including $\tilde{\lambda}$, $\boldsymbol{\beta}$, and $\boldsymbol{u}$. This construction allows the decay parameter $\lambda$ to be inferred jointly with the latent factors $\boldsymbol{\beta}$ and the Gaussian field $\boldsymbol{u}$, fully within the \texttt{inlabru} framework. Crucially, it removes the structural limitation of a fixed observation matrix $\boldsymbol{A}$, effectively extending the range of models that can be fitted using \texttt{inlabru}. Moreover, this mechanism naturally generalizes to more complex settings, such as a time-varying decay parameter $\lambda_t$. Beyond the present model, it opens new avenues for using \texttt{inlabru} in both financial and non-financial applications where non-linear latent relationships are essential. 

In summary, our approach introduces a practical and general framework for incorporating non-linear parameter dependencies inside the observation structure of \texttt{inlabru}, thereby achieving joint Bayesian inference for parameters that were previously inaccessible within the standard \texttt{INLA} methodology.

\section{Yield Curve Forecasting and Economic Value}\label{sec:numerical}
\subsection{Data and models}
\begin{figure}[t!]

\centering{
\includegraphics[width=3in,height=\textheight]{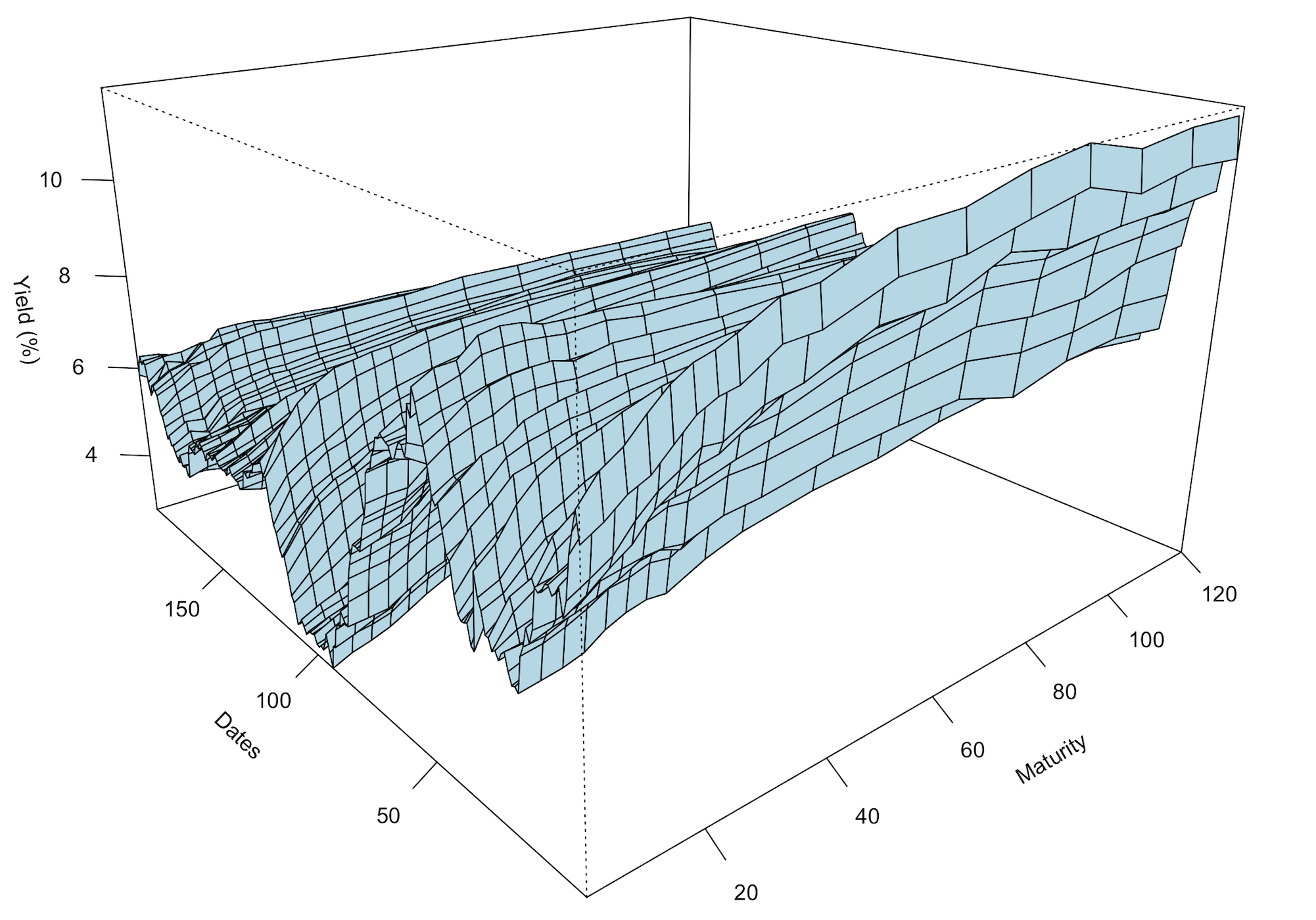}
}
\caption{\label{3d}US Treasury zero-coupon yields. January 1985–December 2000.}
\end{figure}

In this study, we use the database of time series comprising monthly unsmoothed Fama-Bliss US Treasury zero-coupon bond yields from January 1985 to December 2000 taken from the CRSP(Center for Research in Security Prices) government bonds files offered by University of Pennsylvania \url{https://www.sas.upenn.edu/~fdiebold/papers/paper49/FBFITTED.txt}. The dataset covers maturities of 3, 6, 9, 12, 15, 18, 21, 24, 30, 36, 48, 60, 72, 84, 96, 108, and 120 months. This dataset is a classical dataset in forecasting term structure and it is widely used (see, e.g., \cite{diebold2006forecasting} and \cite{xiang2013regime}). A three-dimensional plot of the yield curves analyzed in this study is illustrated in Figure \ref{3d}.

In this paper, we compare several extensions of the DNS model explained in Section~\ref{models} that incorporate residual adjustments using rational SPDE formulations. We consider four specific spatial and spatio-temporal structures: a stationary isotropic field (Stat), a nonstationary field (Nonstat), an anisotropic field (Aniso), and a nonseparable spatio-temporal field (Spatemp) with smoothness parameter $\alpha$ and $\beta = 1$. Because we find that the spatio-temporal field model achieves the best performance, we specifically examine its performance when implemented within a joint estimation framework, where all parameters are estimated simultaneously. We further assess two versions of this joint model, each using a different prior for $\lambda$ (both centered at 0.068): a log-normal prior (Lognormal) with mean 0.068 and coefficient of variation 0.19, and a Gamma prior (Gamma) with mean 0.068 and shape 4. This choice was motivated by empirical evidence that the conventional value of 0.0609 performs suboptimally, while 0.068 emerges as a robust estimate across specifications. The dual prior approach serves as a robustness check, confirming that our results are not overly sensitive to distributional assumptions on the choice of $\lambda$. For comparative purposes, we implement the classical DNS approach as a Bayesian latent Gaussian model. The Bayesian inference of DNS model with no residual adjustment is denoted as BDNS, in line with \cite{diebold2006forecasting}, we treat \(\lambda\) as a fixed constant, assigning it a value of 0.0609, while the two-step estimation approach is our baseline.

\subsection{Out-of Sample Forecasts and Probabilistic Evaluation}
We assess 1-, 6-, and 12-month-ahead forecasting performance for maturities of 3 months and 1, 3, 5, and 10 years. The construction of the forecasts follows the same specifications as used in \cite{diebold2006forecasting}. We conduct estimation and forecasting recursively (in moving windows), beginning with an estimation sample from January 1985 up to the start of the forecast period in January 1995, which extends through December 2000. 

As the core assessment criteria in term structure forecasting, we compare the root mean squared error (RMSE) across models. The baseline DNS model in \cite{diebold2006forecasting} serves as the benchmark, against which we evaluate enhancements from integrating stationary, non-stationary, anisotropic, and nonseparable spatio-temporal SPDE components. 

Based on the forecast performance presented in Table \ref{rmse}, our analysis reveals that most of the models consistently outperform the baseline model. The spatio-temporal model dominates short-term forecasts and short maturities, while the stationary field model excels at longer maturities and longer forecast horizons. The anisotropic model is relatively competitive for intermediate maturities at short horizons. As for the BDNS model, the RMSE decreases in 9 entries out of 15, compared to the baseline approach, and performs especially better in longer horizons. The underperformance of the nonstationary field model is likely caused its overparametrization or insufficient calibration to the residual structure. In general, forecast accuracy deteriorates with horizon length, but our SPDE-driven models mitigate this decay, particularly for long-term rates where DNS rigidities are most significant.

\begin{table}[htbp]
    \centering
    \small
    \setlength{\tabcolsep}{3pt} 
    \caption{Out-of-sample RMSE for yield forecasts at selected maturities, in each row, the best model is highlighted in bold. BDNS denotes the Bayesian DNS model; Stat, Nonstat, Aniso and Spatemp correspond to stationary, non-stationary, anisotropic and spatio-temporal residual models. Lognormal and Gamma refer to spatio-temporal models with $\lambda$ under log-normal and Gamma priors. Baseline corresponds to the classical two-step DNS approach.}
    \begin{subtable}[t!]{\textwidth}
        \centering
        \caption{Out-of-Sample 1-month ahead forecasting}
        \begin{tabular}{l | l l l l l l l l}
        & BDNS & Stat  & Nonstat & Aniso & Spatemp & Lognormal & Gamma & Baseline  \\ \hline
        3 months & 0.149 & 0.150 & 0.189 & 0.149 & 0.143 & \textbf{0.142} & 0.143 & 0.151  \\ 
        1 year & 0.186 & 0.198 & 0.233 & \textbf{0.184} & 0.190 & 0.188 & 0.188 & 0.187  \\ 
        3 years & 0.271 & 0.261 & 0.277 & 0.265 & 0.258 & \textbf{0.256} & \textbf{0.256} & 0.268  \\ 
        5 years & 0.292 & 0.264 & 0.283 & 0.287 & 0.263 & \textbf{0.261} & 0.262 & 0.289  \\ 
        10 years & 0.249 & 0.247 & 0.280 & 0.246 & 0.246 & \textbf{0.245} & \textbf{0.245} & 0.247 \\ 
        \end{tabular}
        \label{tab:rmse1}
    \end{subtable}
    \hfill

    \begin{subtable}[t!]{\textwidth}
        \centering
        \caption{Out-of-Sample 6-month ahead forecasting}
        \begin{tabular}{l | l l l l l l l l}
        & BDNS & Stat  & Nonstat & Aniso & Spatemp & Lognormal & Gamma & Baseline  \\ \hline
        3 months & 0.424 & 0.436 & 0.494 & 0.418 & 0.382 & \textbf{0.379} & 0.381 & 0.428  \\ 
        1 year & 0.572 & 0.586 & 0.627 & 0.557 & 0.558 & 0.557 & \textbf{0.555} & 0.575  \\ 
        3 years & 0.719 & \textbf{0.679} & 0.725 & 0.694 & 0.694 & 0.695 & 0.691 & 0.715  \\ 
        5 years & 0.773 & \textbf{0.689} & 0.741 & 0.747 & 0.719 & 0.722 & 0.717 & 0.768  \\ 
        10 years & 0.708 & \textbf{0.633} & 0.690 & 0.685 & 0.673 & 0.676 & 0.673 & 0.704 \\ 
        \end{tabular}
        \label{tab:rmse2}
    \end{subtable}
    \hfill

    \begin{subtable}[t!]{\textwidth}
        \centering
        \caption{Out-of-Sample 12-month ahead forecasting}
        \begin{tabular}{l | l l l l l l l l}
        & BDNS & Stat  & Nonstat & Aniso & Spatemp & Lognormal & Gamma & Baseline  \\ \hline
        3 months & 0.704 & 0.710 & 0.768 & 0.693 & 0.659 & 0.651 & \textbf{0.648} & 0.717  \\ 
        1 year & 0.783 & 0.783 & 0.819 & 0.768 & 0.777 & 0.775 & \textbf{0.766} & 0.796  \\ 
        3 years & 0.896 & \textbf{0.803} & 0.846 & 0.869 & 0.867 & 0.866 & 0.860 & 0.896  \\ 
        5 years & 0.971 & \textbf{0.828} & 0.878 & 0.941 & 0.913 & 0.911 & 0.905 & 0.973  \\ \
        10 years & 0.960 & \textbf{0.781} & 0.846 & 0.929 & 0.874 & 0.871 & 0.867 & 0.965 \\ 
        \end{tabular}
        \label{tab:rmse3}
    \end{subtable}
    \label{rmse}
\end{table}

While RMSE assesses the models' ability on point forecast accuracy, a comprehensive assessment of Bayesian term structure models demands evaluation of their ability to quantify probablistic forecast uncertainty. To this end, we complement the RMSE analysis with four distributional metrics: the Continuous Ranked Probability Score (CRPS) (\citeay{gneiting2007strictly}), the scaled Continuous Ranked Probability Score (SCRPS) (\citeay{bolin2023local}) and their invariants assessing the tail behavior, weighted CRPS (wCRPS) and scaled wCRPS (swCRPS) (\citeay{olafsdottir2024locally}).

These scores, for a predictive distribution \( P \) and an observation \( y \), are defined as:
\begin{align*}
    \text{CRPS}(P, y) &= \frac{1}{2} \mathbb{E}_{P,P} [ |X - Y| ] - \mathbb{E}_{P} [ |X - y| ], \\
    \text{sCRPS}(P, y) &= - \frac{\mathbb{E}_{P} [ |X - y| ]}{\mathbb{E}_{P,P} [ |X - Y| ]}
    - \frac{1}{2} \log \mathbb{E}_{P,P} [ |X - Y| ], \\
    \text{wCRPS}(P, y) &= \frac{1}{2} \mathbb{E}_{P,P}\left[g_w(X, X')\right] - \mathbb{E}_{P}\left[g_w(X, y)\right], \\
    \text{swCRPS}(P, y) &= -\frac{\mathbb{E}_{P}\left[g_w(X, y)\right]}{\mathbb{E}_{P,P}\left[g_w(X, X')\right]} - \frac{1}{2} \log\left(\mathbb{E}_{P,P}\left[g_w(X, X')\right]\right),
\end{align*}
where \( \mathbb{E}_{P} [g(X)] \) denotes the expected value of a function \( g(X) \) for a random variable $X \sim P$, \( \mathbb{E}_{P,P} [g(X,Y)]\) denotes the expected value of $g(X,Y)$ when $X$ and $Y$ are two independent draws from the predictive distribution $P$, and \(g_w(x, x') = \left|\int_{x'}^{x} w(t) \rm{d}t\right| \) is a kernel function. These scores serve different purposes in evaluating probabilistic forecasts. The CRPS is a standard proper scoring rule that measures the difference between the predicted and empirical cumulative distribution functions. To address its sensitivity to the scale of the forecast distribution, the sCRPS was developed as a locally scale-invariant version, ensuring that score comparisons remain meaningful across forecasts with different inherent variability. The wCRPS generalizes the CRPS by applying a kernel to emphasize specific regions of the distribution, here in the tails. In practice, we target the upper tail where yield rates increase sharply and define a threshold of a 5\% rise from the previous month's yield. Finally, the swCRPS combines these ideas together, creating a locally tail-scale invariant score to ensure fair evaluation of forecasts in these targeted regions, regardless of their scale.

These results, which are shown in Appendix~E, address the limitations of the RMSE, such as insensitivity to tail behavior and overconfidence, while aligning with applications like risk management and derivative pricing, where uncertainty quantification is paramount. 

The out-of-sample forecasting results consistently identify the spatio-temporal SPDE model as the superior specification across virtually all evaluation metrics and horizons. It achieves the lowest RMSE for most maturity-horizon combinations and performs well in uncertainty quantification tests, demonstrating robust performance in both point forecasts and probabilistic calibration.

\subsection{Model Estimates and Residual Diagnostics}
The out-of-sample forecasting results indicate that the nonseparable spatio-temporal SPDE model delivers the strongest overall performance among the competing specifications. To better understand the mechanisms behind these gains, we now examine the corresponding posterior parameter estimates and latent factor dynamics, and compare them to those obtained under the Bayesian DNS (BDNS) model without structured residual dependence.

\begin{figure}[t!p]
  \centering
  \begin{subfigure}[h]{0.3\textwidth}
    \includegraphics[height=1.3\textheight]{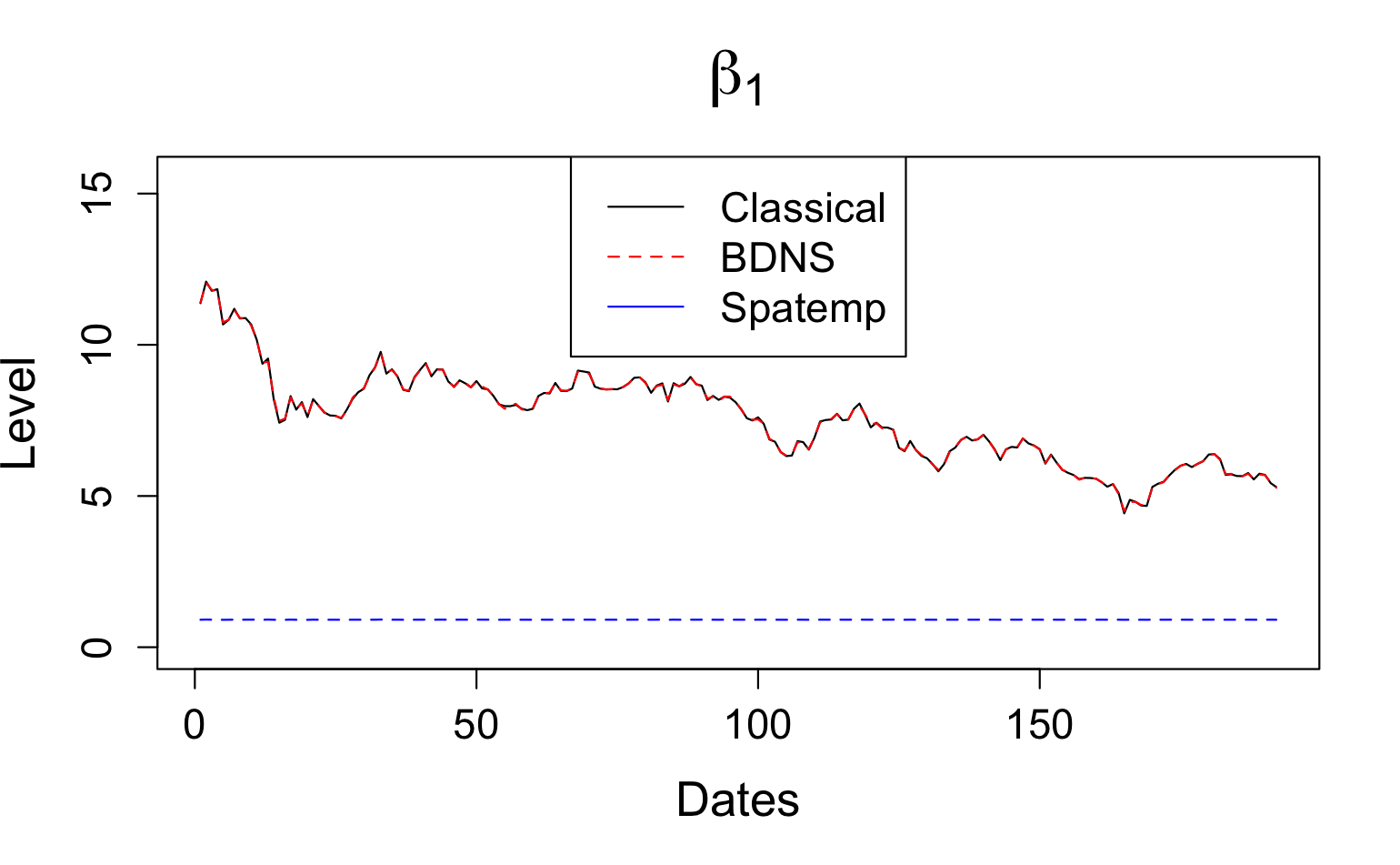}
  \end{subfigure}
  \begin{subfigure}[h]{0.3\textwidth}
    \includegraphics[height=1.3\textheight]{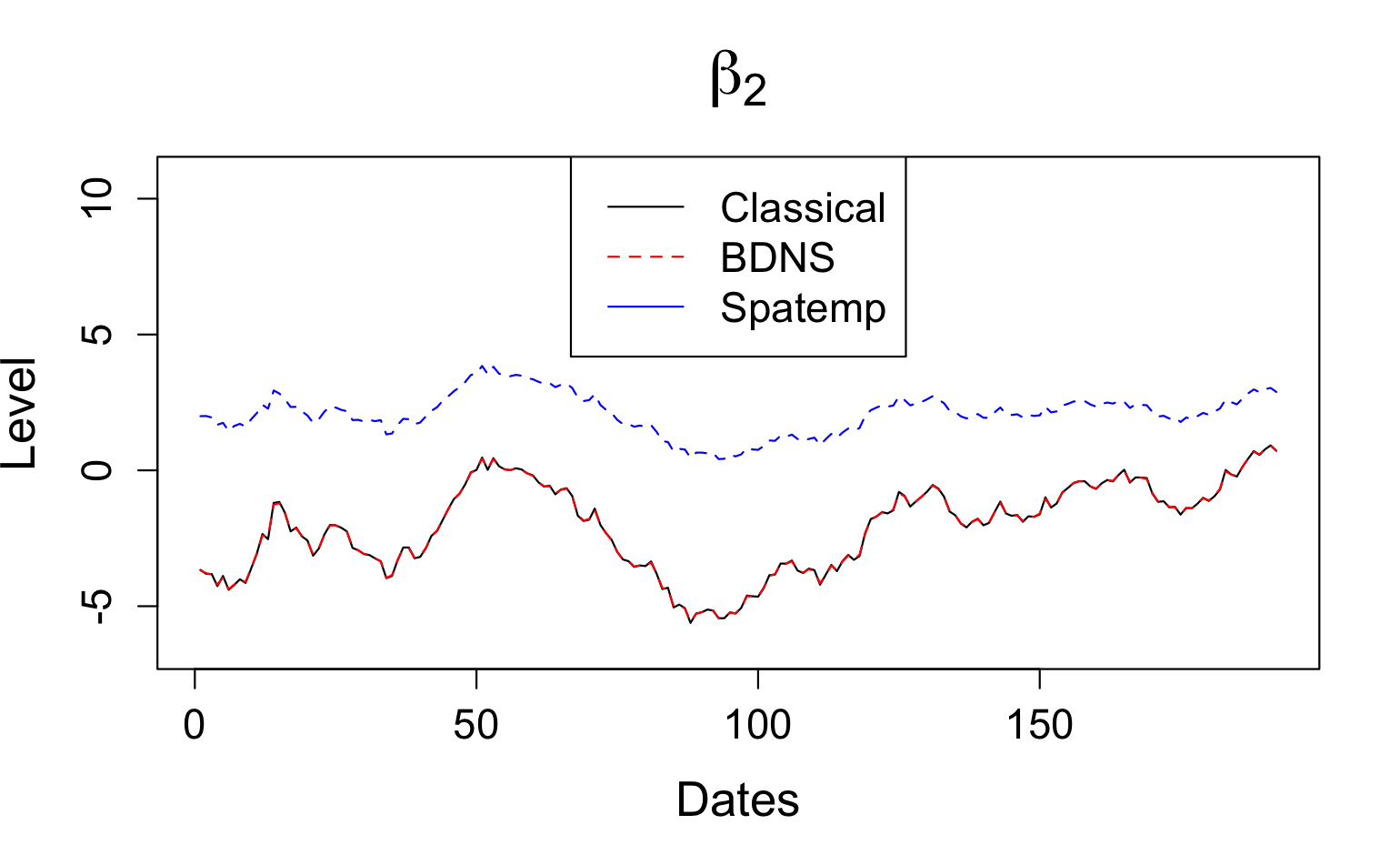}
  \end{subfigure}
  \begin{subfigure}[h]{0.3\textwidth}
    \includegraphics[height=1.3\textheight]{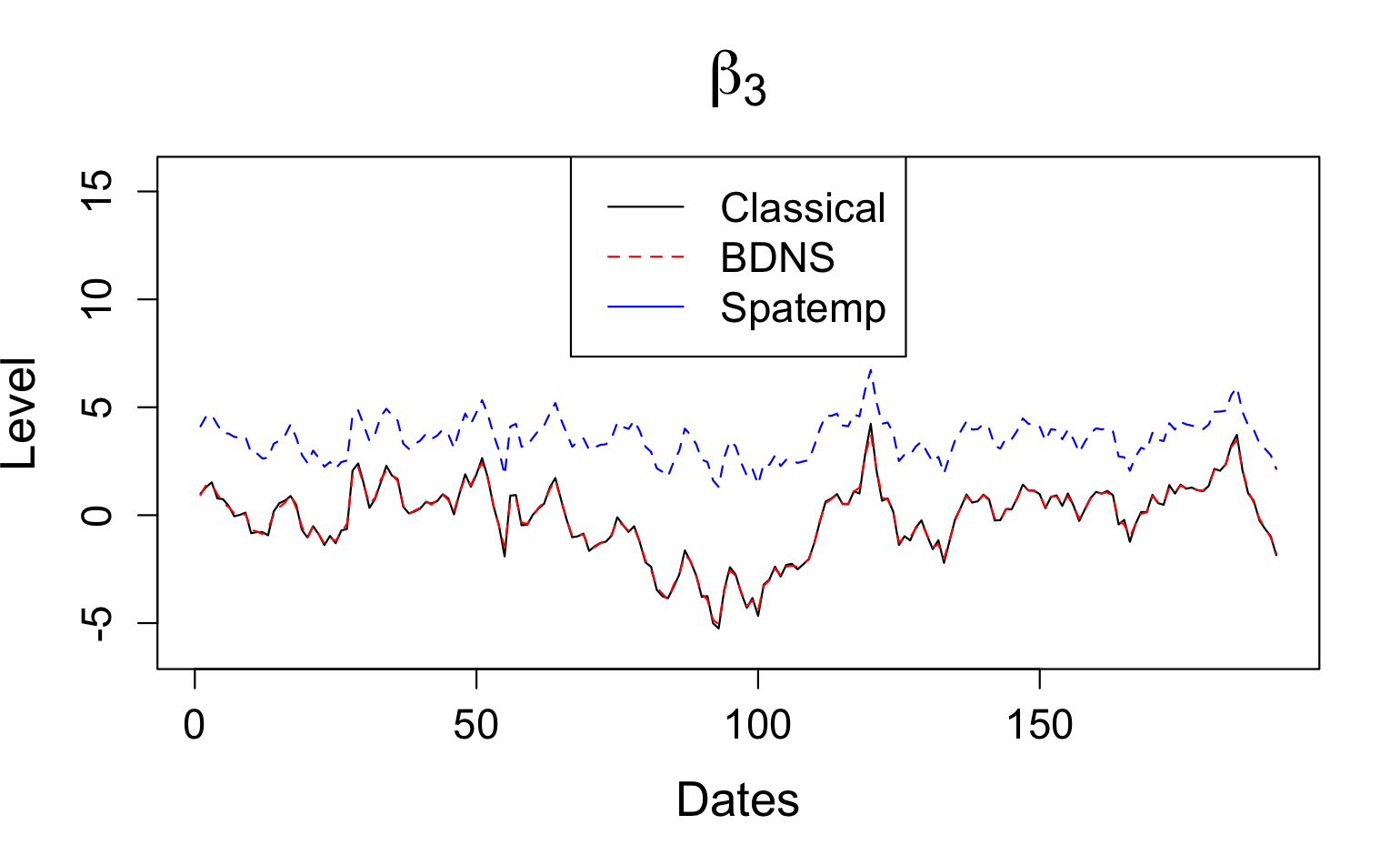}
  \end{subfigure}
  \caption{Posterior mean trajectories of the Nelson–Siegel level, slope, and curvature factors under three estimation approaches: OLS estimates of classical DNS model (black line), BDNS model (red line), and spatio-temporal model (blue line).}
  \label{estimated}
\end{figure}

Figure \ref{estimated} reports the posterior mean trajectories of the Nelson–Siegel level, slope, and curvature factors under three estimation approaches: the classical OLS estimates of \cite{diebold2006forecasting}, the BDNS model, and the proposed spatio-temporal specification. The OLS and BDNS estimates are nearly indistinguishable, confirming that the Bayesian implementation reproduces the classical DNS dynamics. In contrast, introducing the spatio-temporal residual component leads to systematically different factor behavior. The level factor exhibits a markedly flatter trajectory, while the slope and curvature factors retain similar medium-term movements but display reduced short-term variability. Notably, the slope factor changes sign relative to the classical DNS estimates. From an economic perspective, yields at a given point in time typically increase with maturity, implying a positive slope component. Likewise, yield curves often exhibit a hump-shaped profile, with intermediate maturities exceeding both short- and long-term yields, corresponding to a positive curvature component. By explicitly modeling structured cross-maturity dependence in the residual term, the spatio-temporal specification reduces confounding between factor dynamics and residual structure, allowing the slope and curvature factors to adopt economically intuitive signs rather than compensating for unmodeled cross-sectional correlation.

Table \ref{model} presents the posterior summaries of the model parameters for the BDNS and spatio-temporal model, showing the means $\mu_1,\mu_2,\mu_3$ of $\beta_{1t}, \beta_{2t}, \beta_{3t}$, along with estimates for the hyperparameters $\theta_1$ and $\theta_2$ of the these loadings. Following the parametrization outlined in Section~\ref{bayes}, the posterior means for $\tau$ and $\phi$ are obtained by applying the inverse transformations directly to the posterior mean estimates of $\theta_1$ and $\theta_2$. For the spatio-temporal model parameters $\kappa$, $\sigma$, and $\gamma$, the same logarithmic transformation applied to $\tau$ is used ($\xi_1 = \log \kappa, \xi_2 = \log \sigma, \xi_3 = \log \gamma$), ensuring positive values in the model.

\begin{table}[t!]
    \centering
    \small
    \setlength{\tabcolsep}{3pt}
    \caption{Model parameter estimates for BDNS and spatio-temporal models. Each entry shows the BDNS model estimate followed by the spatio-temporal model estimate in parentheses. Parameters marked with dashes (--) are only estimated in the spatio-temporal model.}
    \begin{tabular}{ccccccl}
        \hline
        Parameter & Mean & SD & Parameter & Mean & SD \\
        \hline
        $\mu_1$ & 7.96 (0.91) & 2.50 (28.74) & $\mu_3$ & -0.19 (3.51) & 0.68 (7.90) \\
        $\theta_1$ for $\beta_{1t}$ & 2.48 (9.58) & 0.96 (1.47) & $\theta_1$ for $\beta_{3t}$ & 0.79 (0.92) & 0.38 (0.14) \\
        $\theta_2$ for $\beta_{1t}$ & 4.79 (0.54) & 0.11 (1.04) & $\theta_2$ for $\beta_{3t}$ & 2.47 (1.94) & 0.12 (0.53) \\
        $\mu_2$ & -1.82 (2.23) & 1.83 (8.14) & $\xi_1$ for field & -- (-5.15) & -- (0.09) \\
        $\theta_1$ for $\beta_{2t}$ & 2.37 (3.08) & 0.72 (0.19) & $\xi_2$ for field & -- (-4.04) & -- (0.05) \\
        $\theta_2$ for $\beta_{2t}$ & 4.46 (3.82) & 0.11 (0.68) & $\xi_3$ for field & -- (-1.40) & -- (0.15) \\
        \hline
    \end{tabular}
    \label{model}
\end{table}

Comparing the BDNS model with the spatio-temporal extension reveals systematic differences in parameter estimates that are consistent with the factor trajectories shown in Figure~\ref{estimated}. For the level factor $\beta_{1,t}$, the BDNS model estimates a high mean and strong persistence ($\mu_1 = 7.96$, $\phi_1 = 0.99$), whereas the spatio-temporal model yields a substantially lower mean and weaker persistence ($\mu_1 = 0.91$, $\phi_1 = 0.63$), together with a much higher precision ($\tau_1 = 14535.32$). These estimates imply a tightly constrained and weakly persistent level factor, in line with the flatter trajectory. In contrast to the BDNS specification, the spatio-temporal model yields lower persistence and higher precision for the slope and curvature factors. Once structured dependence across time and maturity is absorbed by the residual field, the remaining factor dynamics become more tightly identified and less persistent, consistent with the faster mean reversion and reduced short-term variability.

This reallocation of explanatory power has important implications for yield curve dynamics. Rather than attributing most movements to parallel level shifts as in classical DNS, the spatio-temporal model distributes variation across slope and curvature factors that capture term-structure dynamics, complemented by a spatial field modeling cross-maturity dependencies. This suggests that apparent level variation in the standard model may actually reflect persistent slope/curvature dynamics and structured cross-sectional dependencies.

\begin{figure}[t!]
  \centering
  \begin{subfigure}[h]{0.8\textwidth}
    \includegraphics[width=0.9\textwidth]{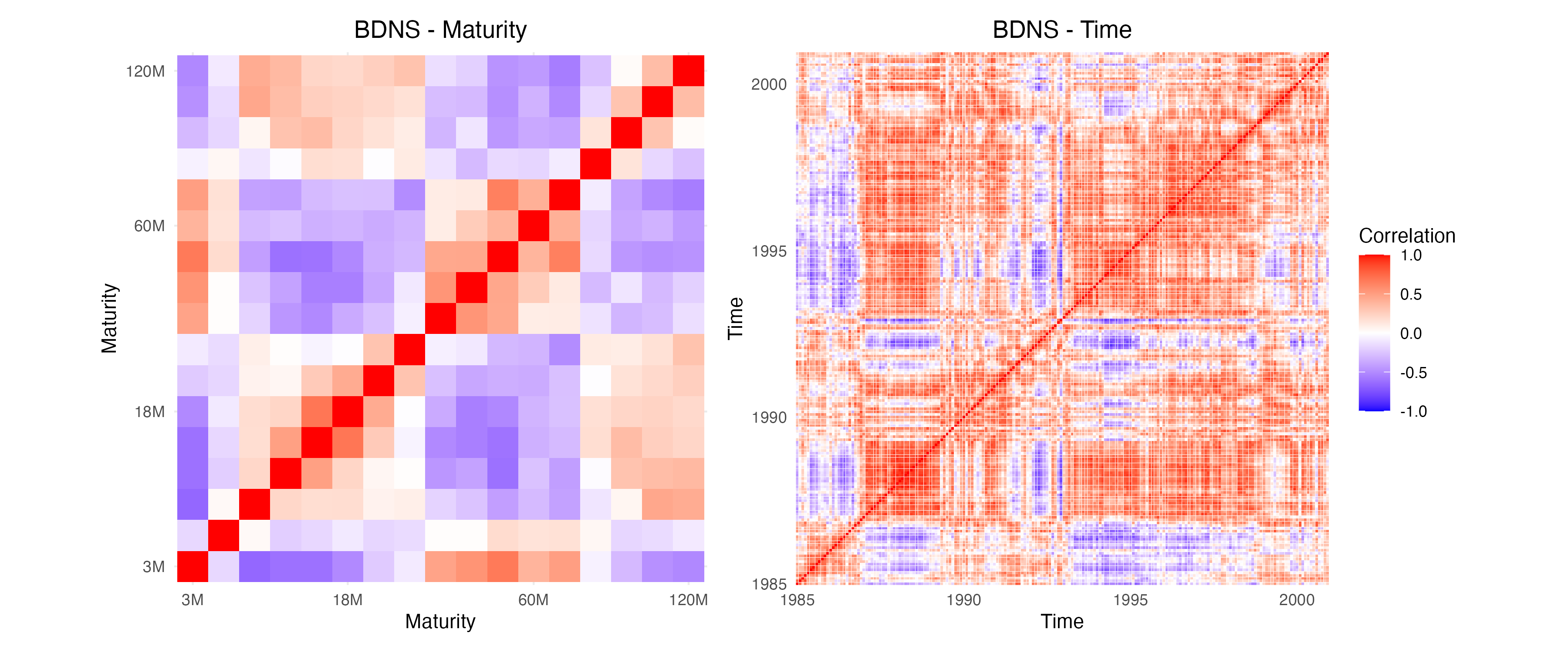}
    \caption{BDNS residual correlations across maturities (left) and across time (right)}
    \label{fig:corr1}
  \end{subfigure}
  \begin{subfigure}[h]{0.8\textwidth}
    \includegraphics[width=0.9\textwidth]{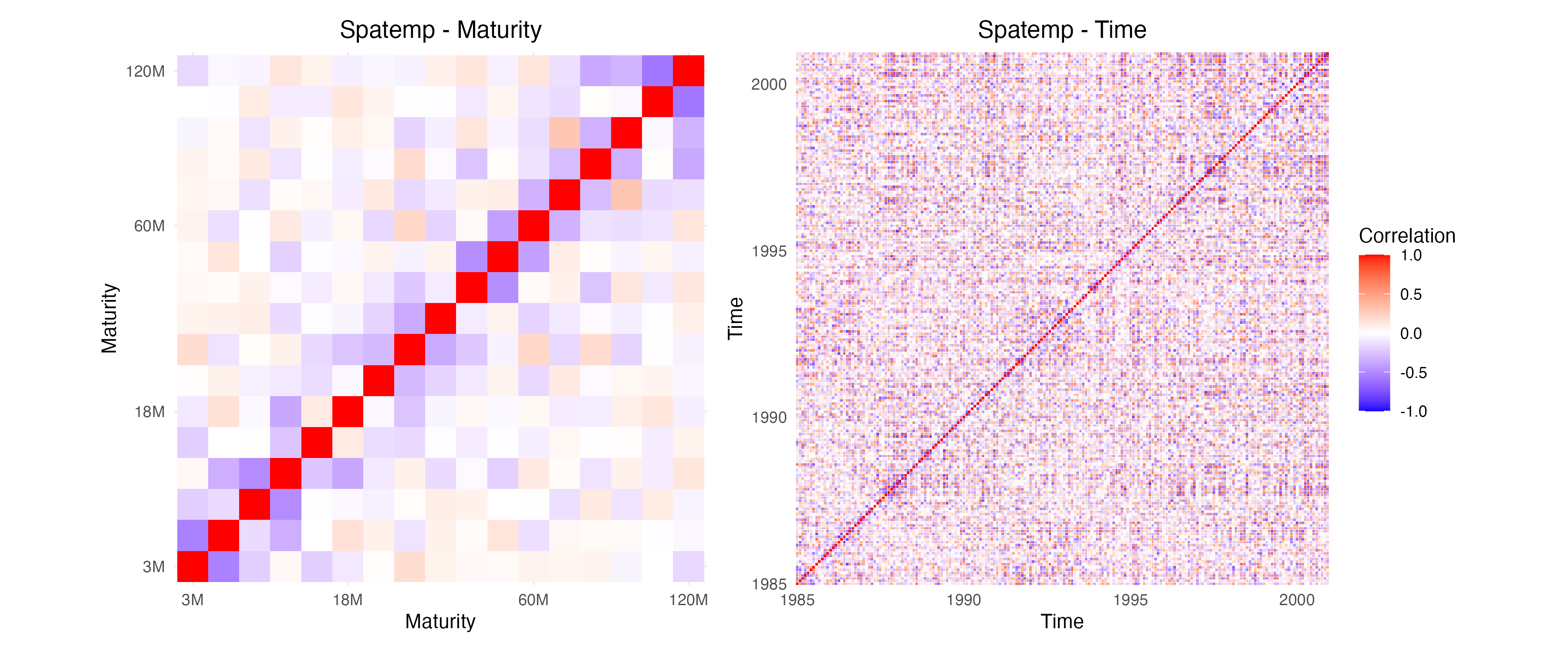}
    \caption{spatemp residual correlations across maturities (left) and across time (right)}
    \label{fig:corr2}
  \end{subfigure}
  \caption{Correlation heatmaps across maturities and across time for residuals of DNS-BDNS model and spatio-temporal model. White indicates zero correlation, while blue and red indicate negative and positive correlations, respectively.}
  \label{corr}
\end{figure}

The spatio-temporal residual field parameters provide further insight. The spatial range parameter ($\kappa = 0.0057$) corresponds to slowly decaying correlations across maturities, indicating long-range cross-sectional dependence in the residuals, while the marginal standard deviation ($\sigma = 0.0175$) suggests moderate amplitude. The temporal evolution parameter ($\gamma = 0.245$) governs the speed at which the spatial field evolves over time, implying persistent but gradually changing residual structures, confirming that the spatio-temporal field captures meaningful dependency patterns that remain unexplained by the latent factors alone.

The differences in parameter estimates are directly linked to model specification in the baseline approach. Figure~\ref{corr} presents correlation heatmaps of residuals for both models. The BDNS residuals exhibit strong correlation patterns across both maturities and time, violating the i.i.d.\ Gaussian noise assumption. Consequently, BDNS parameter estimates should be interpreted with caution, as misspecified residual structure can lead to unreliable inference. In contrast, the spatio-temporal model substantially reduces residual correlation, yielding a structure much closer to white noise. This validates the model assumptions and confirms that the spatio-temporal specification successfully captures the dependence structure that the BDNS model leaves unexplained. Additional residual diagnostics and robustness checks are reported in Appendix~F.

\subsection{Economic Value}
\subsubsection{General framework}
Following the approaches of \citet{carriero2012forecasting} and \citet{xiang2013regime}, we then evaluate the practical usefulness of our competing forecasting models by embedding them into a mean–variance portfolio framework for fixed‐income securities. In line with \citet{markowitz1952portfolio}, investors choose asset weights by balancing expected return against variance.

We assume the investor plans over a $k$‑period horizon and rebalances the portfolio on a monthly basis to maximize the risk-adjusted conditional expected return. To evaluate the economic value of our models, we quantify the performance fee an investor would pay to switch from a benchmark strategy. Determining the optimal time‑varying weights therefore relies on forecasts of the $k$‑step‑ahead conditional mean vector and the corresponding conditional variance–covariance matrix.

\subsubsection{Mean--Variance Portfolio Optimization}

We assume that an investor with an $h$‐period horizon rebalances every $h$ periods. Let $\boldsymbol{w}_t$ denote the $N\times1$ vector of portfolio weights chosen at time $t-h$, $\boldsymbol{Y}(t,m)_{t-h}$ be the vector of $h$‐period‐ahead expected returns, and $\boldsymbol{\Sigma}_Y$ be the return covariance matrix. We solve
\begin{equation}
\begin{aligned}
\min_{\boldsymbol{w}_t}\quad & \boldsymbol{w}_t^\top \,\boldsymbol{\Sigma}_Y\, \boldsymbol{w}_t \;-\;\frac{1}{\delta}\,\boldsymbol{w}_t^\top \,\boldsymbol{Y}(t,m)_{t-h}\\
\text{s.t.}\quad & \boldsymbol{w}_t^\top \,\boldsymbol{1}=1,
\end{aligned}
\end{equation}
where $\delta$ is the investor’s risk‐aversion parameter.

\subsubsection{Performance Fee}
To connect the returns generated by dynamically rebalanced portfolios to expected utility, we adopt the \cite{west1993utility} framework. Under the assumption of a quadratic utility function, the investor’s realized utility takes the form:
\begin{equation}
    U(W_{t}) = W_{t} - \frac{\zeta}{2} W_{t}^2 = W_{t-h}R_{t} -  \frac{\zeta}{2}W_{t-h}^2R_{t}^2, \quad t=1,\ldots,T,
\end{equation}
where $W_{t-h}$ is the investor’s wealth at $t-h$, $\zeta$ is a risk-preference parameter of utility and $R_{t} = 1 + \boldsymbol{w}_t^\top \,\boldsymbol{Y}(t,m)_{t-h}$ is the period $t$ gross return on his or her portfolio.

Assuming the investor's relative risk aversion (RRA) remains constant and is defined by $\bar{\delta} = \zeta W_t / (1- \zeta W_t)$, it can be shown that the expected utility derived from an initial wealth level equals the average utility $\bar{U}$, which can be computed as
\begin{align}
    \bar{U} = W_0\sum_{t=1}^{T-1} \left(R_{h} - \frac{\bar{\delta}}{2(1+\bar{\delta})}R_{h}^{2} \right),
    \label{wealth}
\end{align} for an initial wealth $W_0$. For simplicity, we here set $W_0 = 1$ and $\bar{\delta} = 1$.

The economic value is then measured by the maximum fee $F$ an investor would pay to adopt the residual-adjusted models over a benchmark. Here, we pick the Bayesian extension of the DNS model as the benchmark to check the performance of our residual model. Let $R_{h}^{\text{Model}}$ denote gross portfolio return of our proposed models including the stationary, anisotropic and spatio-temporal model, and $R_{h}^{\text{Benchmark}}$ denote the gross return of a dynamically rebalanced portfolio constructed using the BDNS model. This portfolio serves as the benchmark of our model performances. The performance fee $F$ solves:
\begin{align*}
    \sum_{t=1}^\top \left[ \left(R_{h}^{\text{Model}} - F\right) -\frac{1}{4} \left(R_{h}^{\text{Model}} - F\right)^2 \right] =
    \sum_{t=1}^\top \left[ R_{h}^{\text{Benchmark}} -\frac{1}{4} \left(R_{h}^{\text{Benchmark}}\right)^2 \right].
\end{align*}
For the economic‐value calculation, we model the investor’s portfolio as comprising two instruments: a 10‑year Treasury bond and a short‑term Treasury bond with maturity $m$.  The 10‑year bond is treated as risk‐free since its yield is known in advance, whereas the rolling strategy in the short‑term bond entails risk due to uncertain future yields.

\begin{table}[t!]
    \centering
    \small 
    \setlength{\tabcolsep}{2pt} 
    \caption{Economic value under alternative risk preference assumptions. Higher values indicate greater portfolio returns under fixed volatility constraints. For each risk-preference $\zeta$ (row blocks) and each maturity (columns), the best model is highlighted in bold.}
    \begin{tabular}{l | l l  l  l  l}
    \hline
        Risk-preference & Model & 3-month & 12-month & 36-month & 60-month \\ \hline
        $\zeta$ = 4 & Stat & 3.67 & 2.84 & \textbf{3.73} & \textbf{7.15} \\ 
        & Aniso & 1.40 & 2.80 & 2.65 & 2.64 \\
        & Spatemp & 8.52 & \textbf{5.58} & 3.49 & 6.82 \\ 
        & Lognormal & \textbf{8.64} & 4.87 & 3.43 & 7.11 \\
        & Gamma & 8.09 & 4.60 & 3.38 & 7.04 \\  \hline
        $\zeta$ = 2 & Stat & 4.99 & 3.81 & 5.97 & 12.64  \\
        & Aniso & 1.79 & 3.99 & 4.30 & 4.47 \\
        & Spatemp & 13.61 & \textbf{9.53} & \textbf{6.49} & 12.75 \\
        & Lognormal & \textbf{13.69} & 8.40 & 6.41 & \textbf{13.34} \\
        & Gamma & 12.98 & 8.04 & 6.35 & 13.24 \\ \hline
        $\zeta$ = 1 & Stat & 5.81 & 4.71 & 9.32 & 21.87 \\
        & Aniso & 2.07 & 5.15 & 6.78 & 7.54 \\
        & Spatemp & 17.53 & \textbf{13.45} & \textbf{10.98} & 22.73 \\
        & Lognormal & \textbf{17.60} & 11.91 & 10.87 & \textbf{23.81} \\ 
        & Gamma & 16.77 & 11.47 & 10.78 & 23.64 \\
    \end{tabular}
    \label{fee}
\end{table}

We calculate the performance fee for each portfolio of the 10-year bond (risk-free rate) and the short-term bond. Table \ref{fee} presents the percentage ratio of performance fees with different risk aversion preferences to the gross return of the benchmark model. Only stationary, anisotropic, spatio-temporal, and two joint model variants are considered, as they have demonstrated consistently high prediction accuracy in our analysis. Because the portfolio is monthly rebalanced, the rebalance is based on the 1-month-ahead forecasts. The findings indicate that the performance‐based fees associated with moving from the benchmark model to the residual-adjusted models are generally substantial. For example, with a medium risk aversion preference ($\zeta = 2$), the portfolio selected by the spatio-temporal model will yield an extra 13\% in the average 3-month return compared to the benchmark model.

\section{Conclusion}\label{sec:conclusion}
We propose a flexible and computationally efficient extension of the Dynamic Nelson–Siegel framework that explicitly accounts for residual dependence across both calendar time and maturity. By modeling the unexplained component of yields as a Gaussian random field defined through stochastic partial differential equations, we move beyond the conventional assumption of conditionally independent measurement errors while preserving the interpretability of the Nelson–Siegel factors. Across a range of specifications, including stationary, non-stationary, anisotropic, and nonseparable spatio-temporal formulations, the SPDE-based residual models deliver systematic improvements in out-of-sample forecasting performance relative to classical DNS benchmarks. These gains are particularly significant for short-horizon forecasts and short maturities. Evaluation using proper scoring rules further indicates that the proposed framework improves uncertainty quantification.

The nonseparable spatio-temporal specification emerges as the most robust and empirically successful model. Parameter estimates and latent factor trajectories reveals that introducing structured residual dependence alters how variation in the yield curve is decomposed between the Nelson–Siegel factors and the residual component. In particular, cross-maturity movements that are traditionally attributed to the level factor are captured by the spatio-temporal field once long-range dependence across maturities is explicitly modeled. 

Methodologically, our approach demonstrates how SPDE-based Gaussian random fields can be seamlessly integrated into latent factor term-structure models within a Bayesian framework. By leveraging INLA and the rational SPDE construction, we allow flexible control over smoothness, range, anisotropy, and temporal evolution. In addition, the use of \texttt{inlabru} enables joint estimation of the Nelson–Siegel decay parameter together with latent factors and residual fields, providing coherent uncertainty quantification for all model components.

The economic relevance of these improvements is confirmed through a mean–variance portfolio exercise, where forecasts from the SPDE-augmented models translate into economically meaningful utility gains relative to a Bayesian DNS benchmark. This highlights that improved statistical modeling of residual dependence can have direct implications for portfolio allocation and risk management decisions.

Beyond forecasting, the proposed SPDE–DNS framework offers a natural tool for yield-curve interpolation and smoothing across maturities, while maintaining economic interpretability. Future research directions include extending the framework to multivariate yield systems, incorporating macroeconomic covariates, and allowing for stochastic volatility within the SPDE component. Overall, combining Nelson–Siegel factor dynamics with SPDE-driven residual modeling provides a coherent and practical framework for flexible, interpretable, and scalable term-structure analysis.

\section{Disclosure statement}\label{disclosure-statement}

The authors declare that they have no conflicts of interest.

\section{Data Availability Statement}\label{data-availability-statement}

The dataset supporting the findings of this study, is publicly available at the following URL:\\
\url{https://www.sas.upenn.edu/~fdiebold/papers/paper49/FBFITTED.txt}

\bibliography{bibliography.bib}

\renewcommand{\thesection}{A}
\section{Supplementary Material}
\subsection{Gaussian White Noise and Weak Solutions to SPDEs}\label{whitenoise}

This section reviews the concept of Gaussian white noise and the weak formulation of stochastic partial differential equations (SPDEs) that define Matérn Gaussian fields.

Gaussian white noise, denoted \( \mathcal{W}(\boldsymbol{s}) \), is a generalized stochastic process defined as a random linear functional on the Hilbert space \( L^2(D) \).
For any test function \( \varphi \in L^2(D) \),
\[
  \mathcal{W}(\varphi) = \int_D \varphi(\boldsymbol{s})\, \mathcal{W}(\mathrm{d}\boldsymbol{s})
\]
is a Gaussian random variable with mean zero and the covariance between $\mathcal{W}(\varphi_1)$ and $\mathcal{W}(\varphi_2)$ for $\varphi_1, \varphi_2 \in L^2(D)$ is 
\[
  \mathbb{E}\big[\mathcal{W}(\varphi_1)\mathcal{W}(\varphi_2)\big]
  = (\varphi_1, \varphi_2)_{L^2(D)}
  = \int_D \varphi_1(\boldsymbol{s})\varphi_2(\boldsymbol{s})\,\mathrm{d}\boldsymbol{s}.
\]
Hence, Gaussian white noise has unit variance per unit measure and independent increments over disjoint regions. That is, if the supports of $\varphi_1$ and $\varphi_2$ are disjoint, i.e.,$ \operatorname{supp}(\varphi_1) \cap \operatorname{supp}(\varphi_2) = \emptyset,$ then $\mathbb{E}\big[\mathcal{W}(\varphi_1) \mathcal{W}(\varphi_2)\big] = 0.$

Consider the stochastic partial differential equation
\begin{equation}
    (\kappa^2 - \Delta)^{\alpha/2} u(\cdot) = \mathcal{W}(\cdot), 
  \quad  \text{on} \ D \subset \mathbb{R}^d,
  \label{standardspde}
\end{equation}
where \( \kappa > 0 \) controls the correlation range and \( \alpha > 0 \) determines the smoothness of the field.
Since \( \mathcal{W} \) is a distribution rather than a pointwise-defined function, the SPDE is understood in the weak sense.
That is, we seek a random field \( u \) such that
\[
  a_L(u, \varphi) = (\mathcal{W}, \varphi)
  \quad \forall \varphi \in H^{\alpha/2}(D),
\]
where \( a_L(\cdot,\cdot) \) is the bilinear form associated with the operator \( L = (\kappa^2 - \Delta)^{\alpha/2} \).

To use this model, we employ the finite element method.
The Gaussian white noise \( \mathcal{W} \) on \( L^2(D) \) is discretized in the finite element space \( V_h \subset H^1(D) \) as
\[
  \mathcal{W}_h = \sum_{j=1}^{n_h} \xi_j e_{j,h},
\]
where \( \{\xi_j\}_{j=1}^{n_h} \) are independent standard Gaussian random variables and \( \{e_{j,h}\}_{j=1}^{n_h} \) form an orthonormal basis of \( V_h \).
The discrete SPDE then reads $L_h^{\alpha/2} u_h = \mathcal{W}_h,$
whose solution can be expressed as
\[
  u_h(\boldsymbol{s}) = \sum_{j=1}^{n_h} w_j \varphi_j(\boldsymbol{s}),
\]
with stochastic weights \( \boldsymbol{w} = (w_1, \ldots, w_{n_h})^\top \).
The vector \( \boldsymbol{w} \) follows a Gaussian Markov random field with precision matrix \( \boldsymbol{Q}(\kappa, \alpha) = L_h^{\alpha} \).
For instance, when \( \alpha = 2 \),
\[
  \boldsymbol{Q}(\kappa, 2) = \kappa^4 \boldsymbol{C} + 2\kappa^2 \boldsymbol{G} + \boldsymbol{G}\boldsymbol{C}^{-1}\boldsymbol{G},
\]
where
\[
  C_{ij} = \int_D \varphi_i(\boldsymbol{s})\varphi_j(\boldsymbol{s})\,\mathrm{d}\boldsymbol{s},
  \qquad 
  G_{ij} = \int_D \nabla\varphi_i(\boldsymbol{s}) \cdot \nabla\varphi_j(\boldsymbol{s})\,\mathrm{d}\boldsymbol{s},
\]
$\boldsymbol{C}$ and $\boldsymbol{G}$ are the mass and stiffness matrices, respectively. This construction yields a sparse precision matrix, providing a computationally efficient approximation to the continuous Gaussian field while preserving its covariance structure.

The SPDE framework thus establishes a rigorous connection between continuous Gaussian fields and discrete Gaussian Markov random fields. Gaussian white noise introduces spatial randomness at infinitesimal scales, while the weak formulation ensures that \( u(\boldsymbol{s}) \) inherits smoothness and correlation properties from the elliptic operator. This link underlies modern SPDE-based spatial modeling and justifies the use of FEM and INLA for scalable Bayesian inference.

\subsection{Finite Element Method} \label{FEM}
Let $\mathcal{D} \subset \mathbb{R}^d$ be a bounded domain. We begin by defining the relevant function spaces.

The space of square-integrable functions on $\mathcal{D}$ is denoted by $L_2(\mathcal{D})$ and equipped with the inner product:
\[
(\phi,\psi)_{L_2(\mathcal{D})} = \int_{\mathcal{D}} \phi(\boldsymbol{x})\psi(\boldsymbol{x})\rm{d}\boldsymbol{x}.
\]

For integer orders $k \in \mathbb{N}$, the Sobolev space $H^k(\mathcal{D})$ consists of functions whose weak derivatives up to order $k$ are in $L_2(\mathcal{D})$:
\[
H^k(\mathcal{D}) = \left\{w \in L_2(\mathcal{D}) : D^{\gamma}w \in L_2(\mathcal{D}), \forall \gamma \in \mathbb{N}^d, |\gamma| \leq k\right\},
\]
Here, $D^{\gamma}$ denotes the weak derivative associated with the multi-index 
$\gamma = (\gamma_1,\dots,\gamma_d) \in \mathbb{N}^d$, where the derivative order is 
$|\gamma| = \gamma_1 + \cdots + \gamma_d$ and
\[
D^{\gamma} w = 
\frac{\partial^{|\gamma|} w}{\partial x_1^{\gamma_1}\cdots \partial x_d^{\gamma_d}}.
\]
Thus, requiring $D^{\gamma} w \in L_2(\mathcal{D})$ for all $|\gamma| \le k$ ensures that all weak partial derivatives of total order up to $k$ are square-integrable. 
The Sobolev inner product is then given by
\[
(u,v)_{H^k(\mathcal{D})} = \sum_{\gamma \in \mathbb{N}^d: |\gamma| \leq k} (D^{\gamma}u, D^{\gamma}v)_{L_2(\mathcal{D})}.
\]

For fractional orders $\sigma > 0$, $\sigma \notin \mathbb{N}$, the fractional Sobolev space $H^\sigma(\mathcal{D})$ is defined via real interpolation (\citeay{chandler2015interpolation}):
\[
H^{\sigma}(\mathcal{D}) = 
\begin{cases}
[L_2(\mathcal{D}), H^1(\mathcal{D})]_{\sigma}, & \text{for } 0 < \sigma < 1, \\
[H^1(\mathcal{D}), H^2(\mathcal{D})]_{\sigma-1}, & \text{for } 1 < \sigma < 2.
\end{cases}
\]

The differential operator in the previous SPDEs are assumed to be the second-order differential operator in divergence form:
\[
Lu = -\nabla \cdot (\boldsymbol{H}\nabla u) + \kappa^2 u,
\]
where $\boldsymbol{H}:\mathcal{D}\rightarrow\mathbb{R}^{d\times d}$ is a symmetric, Lipschitz continuous and uniformly positive definite function, and $\kappa:\mathcal{D}\rightarrow\mathbb{R}$ is an essentially bounded function. Under Neumann boundary conditions, we additionally require that $\kappa > 0$.

This operator induces a continuous and coercive bilinear form on $V$, where $V = H^1_0(\mathcal{D})$ for Dirichlet boundary conditions or $V = H^1(\mathcal{D})$ for Neumann boundary conditions:
\[
a_L(v,u) = (\boldsymbol{H}\nabla u, \nabla v)_{L_2(\mathcal{D})} + (\kappa^2 u, v)_{L_2(\mathcal{D})}, \quad u,v \in V.
\]

For a bounded domain $\mathcal{D} \subset \mathbb{R}^2$ (representing our spatial domain of interest), we begin by constructing a triangulation $\mathcal{T}_h$ with mesh width parameter $h > 0$ controlling the element sizes. The finite-dimensional approximation space $V_h \subset V$ is spanned by continuous piecewise linear basis functions $\{\varphi_j\}_{j=1}^{n_h}$, where each $\varphi_j$ equals 1 at node $j$ and decreases linearly to 0 at adjacent nodes, forming the characteristic hat functions.

In two dimensions, on a triangular element with vertices $(x_1,y_1), (x_2,y_2), (x_3,y_3)$, the basis function associated with vertex 1 is given by the barycentric coordinate:
\[
\varphi_1(x,y) = \frac{(y_2-y_3)(x-x_3) + (x_3-x_2)(y-y_3)}{(y_2-y_3)(x_1-x_3) + (x_3-x_2)(y_1-y_3)}
\]
with similar expressions for $\varphi_2(x,y)$ and $\varphi_3(x,y)$.

The FEM discretization of the differential operator $L$ is determined by restricting the bilinear form $a_L(\cdot,\cdot)$ to the finite element space $V_h \times V_h$. This yields the discrete operator $L_h: V_h \to V_h$ defined through the relation:
\[
(L_h \phi_h, \psi_h)_{L_2(\mathcal{D})} = a_L(\phi_h, \psi_h), \quad \forall \phi_h, \psi_h \in V_h.
\]

\begin{figure}[t!]
    \centering
    \includegraphics[width=0.5\linewidth]{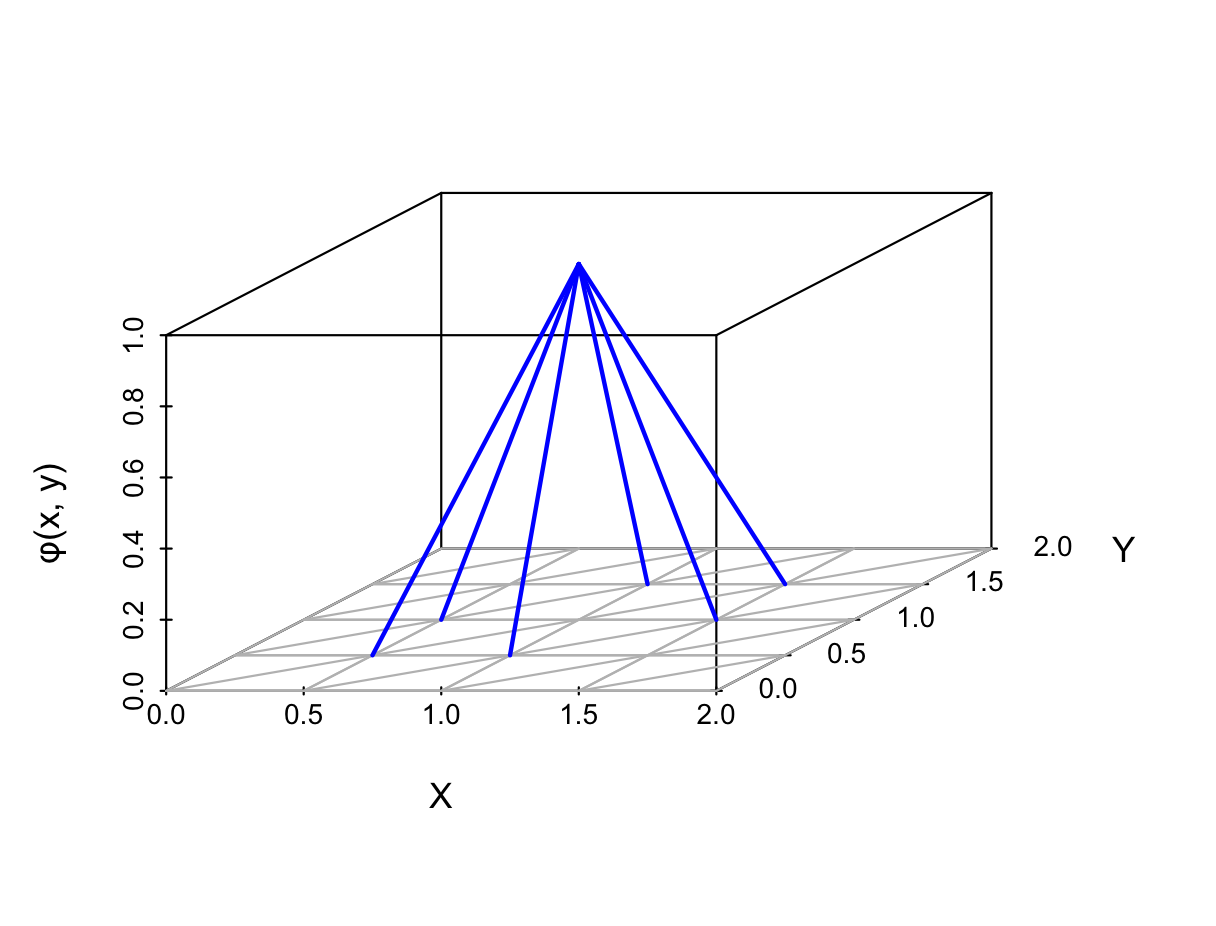}
    \caption{Linear Hat Basis Function in a 2D Triangular Mesh}
    \label{basis}
\end{figure}

Figure \ref{basis} shows a linear hat basis function defined on a triangular mesh with mesh size $h = 0.5$, illustrating the piecewise linear nature of the basis functions used in the FEM approximation.

\subsection{Rational Approximation} \label{rationalapprox}
To bring more flexibility to the estimation of the smoothness parameter, \cite{bolin2024covariance} proposed a method to approximate the covariance operator $L^{-\alpha} = (\kappa^2 - \Delta)^{-\alpha}$ of the random field directly by a finite element method combined with a rational approximation of the fractional power. 

By discretizing the domain using FEM and define $L_h$ as the FEM approximation of $L$. The fractional power is then approximated by splitting:
\[
L_h^{-\alpha} = L_h^{-\lfloor \alpha \rfloor} \cdot L_h^{-\{\alpha\}},
\]
where $\lfloor \alpha \rfloor$ is the integer part and $\{\alpha\} = \alpha - \lfloor \alpha \rfloor$ is the fractional part.

The fractional component $L_h^{-\{\alpha\}}$ is approximated by using a rational function:
\[
L_h^{-\{\alpha\}} \approx p(L_h^{-1}) \cdot q(L_h^{-1})^{-1},
\]
where $p(x) = \sum_{i=0}^{m} a_i x^{m - i}$ and $q(x) = \sum_{j=0}^{m} b_j x^{m - j}$ are polynomials of degree $m$ obtained via best rational approximation developed by \cite{hofreither2021algorithm}.

Hence, the full approximation becomes:
\[
L_h^{-\alpha} \approx L_{h,m}^{-\alpha} := L_h^{-\lfloor \alpha \rfloor} \cdot p(L_h^{-1}) \cdot q(L_h^{-1})^{-1}.
\]

This operator is then used to approximate the covariance function:
\[
\varrho_{h,m}^{\beta}(x, y) = \sum_{j=1}^{n_h} \lambda_{j,h}^{-\lfloor \alpha \rfloor} \cdot p(\lambda_{j,h}^{-1}) \cdot q(\lambda_{j,h}^{-1})^{-1} \cdot e_{j,h}(x) e_{j,h}(y),
\]
where $\lambda_{j,h}$ and $e_{j,h}$ are the eigenvalues and eigenfunctions of $L_h$.

The complete error bound combining FEM discretization and rational approximation of the true covariance function is:
\[
\|\varrho_{h,m}^{\beta} - \varrho^\beta\|_{L_2(\mathcal{D}\times\mathcal{D})} \lesssim h^{\min\{4\beta-1-\varepsilon,2\}} + h^{-1} e^{-2\pi\sqrt{\{2\beta\}m}}
\]
where $m$ is the order of rational approximation and $\varrho$ is the covariance function of the solution of \eqref{standardspde}. The optimal choice $m = 2$ provides a practical balance between accuracy and computational efficiency for our yield curve application.

This bypasses the limit of smoothness parameter and enables flexible and fast inference for arbitrary smoothness parameters.

\subsection{Space-time processes} \label{stprocess}
Let $H$ be a separable Hilbert space, e.g., $H = L^2(M)$ for a spatial domain $M$, and let $Q \colon H \to H$ be a symmetric, positive semidefinite operator. 
A $Q$-Wiener process $\{ W_Q(t) \}_{t \ge 0}$ on $H$ is formally defined by
\[
    W_Q(t) = \sum_{j=1}^{\infty} \sqrt{\lambda_j}\, \beta_j(t)\, e_j,
\]
where $\{e_j\}$ are the eigenfunctions of $Q$ with eigenvalues $\{\lambda_j\}$, and $\{\beta_j(t)\}$ are independent standard Brownian motions. 

If $Q$ is trace-class ($\sum_j \lambda_j < \infty$), $W_Q(t)$ is a proper $H$-valued random variable with covariance $ \mathbb{E}[\, W_Q(t) \otimes W_Q(s) \,] = (t \wedge s)\, Q.$
The cylindrical Wiener process is the formal limit $W_I(t)$ obtained by taking $Q = I$, the identity operator, which is not trace‐class on infinite‐dimensional Hilbert spaces. Hence the process is not H-valued in the usual sense, just “cylindrical.”

Consider a linear stochastic partial differential equation
\[
    \mathrm{d}u(t) = A u(t)\, \mathrm{d}t + \, \mathrm{d}W_Q(t), 
    \quad t \ge 0, \quad u(0) = u_0,
\]
where $A$ generates a $C_0$-semigroup $\{S(t)\}_{t\ge 0}$ on $H$ and $W_Q(t)$ is a $Q$-Wiener process as defined above.  In our case, $A = \gamma (\kappa^2 + \kappa^{d/2} \rho \cdot \nabla_m - \Delta_m)^{\alpha}.$

\subsection{Uncertainty Quantification} \label{uncertainty}
Based on the RMSE-driven insights, this analysis examines whether the same models excel probabilistically or if new trade-offs emerge, thereby offering a holistic view of the DNS-SPDE framework’s forecasting utility. Tables \ref{crps}, \ref{scrps}, \ref{wcrps}, and \ref{swcrps} show the utility in forecasting 1 month ahead (1M), 6 months ahead (6M), and 12 months ahead (12M) yield rates. The thresholds for wCRPS and SwCRPS are set at a 5\% increase over the previous yield value, to examine the models’ ability to capture extreme movements. For example, if the previous month’s yield is 1, the threshold is 1.05. For reference, the models considered are BDNS model (AR(1)) as baseline, BDNS model against four residual adjustment specifications: stationary field (Stat), nonstationary field (Nonstat), anisotropic field (Aniso), and spatiotemporal process (Spatemp) and the joint model under two priors for $\lambda$ centered at 0.068: log-normal (Lognormal) with mean 0.068 and CV 0.19, and gamma (Gamma) with shape 4 and mean 0.068.

\begin{table}[t!]
    \centering
    \caption{Continuous Ranked Probability Score (CRPS), measuring overall probabilistic forecast accuracy based on the full predictive distribution. The stationary model achieves the lowest CRPS for long maturities and long forecast horizons, while the spatiotemporal model with Gamma and lognormal priors performs best for short maturities and short horizons. For each row, the best model is highlighted in bold.}
    \begin{tabular}{l l l l l l l l l }
    \hline
        Horizons & Maturity & AR(1) & Stat & Nonstat & Aniso & Spatemp & Lognormal & Gamma \\ \hline
        1M & 3 months & 0.122 & 0.123 & 0.132 & 0.123 & 0.106 & \textbf{0.106} & 0.106 \\ 
        ~ & 1 year & 0.131 & 0.135 & 0.146 & 0.130 & 0.122 & 0.122 & \textbf{0.121} \\ 
        ~ & 3 years & 0.159 & 0.155 & 0.163 & 0.157 & 0.149 & 0.149	& \textbf{0.148} \\ 
        ~ & 5 years & 0.167 & 0.153 & 0.163 & 0.164 & 0.149 & \textbf{0.149}	& 0.149 \\ 
        ~ & 10 years & 0.143 & 0.142 & 0.157 & 0.142 & 0.139 & \textbf{0.138} & 0.138 \\ \hline
        6M & 3 months & 0.308 & 0.309 & 0.325 & 0.307 & 0.278 & 0.278 & \textbf{0.276}\\ 
        ~ & 1 year & 0.355 & 0.359 & 0.375 & 0.349 & 0.340 & 0.341 & \textbf{0.340}\\ 
        ~ & 3 years & 0.418 & \textbf{0.399} & 0.424 & 0.405 & 0.405 & 0.406 & 0.403\\ 
        ~ & 5 years & 0.446 & \textbf{0.403} & 0.434 & 0.432 & 0.418 & 0.419	& 0.416\\ 
        ~ & 10 years & 0.410 & \textbf{0.370} & 0.406 & 0.398 & 0.391 & 0.393 & 0.391\\ \hline
        12M & 3 months & 0.462 & 0.457 & 0.463 & 0.460 & 0.433 & 0.429 & \textbf{0.427}\\
        ~ & 1 year & 0.482 & 0.473 & 0.476 & 0.477 & 0.471 & 0.469 & \textbf{0.466}\\ 
        ~ & 3 years & 0.521 & \textbf{0.475} & 0.486 & 0.508 & 0.507 & 0.506 & 0.504\\ 
        ~ & 5 years & 0.558 & \textbf{0.483} & 0.503 & 0.542 & 0.528 & 0.527	& 0.525\\ 
        ~ & 10 years & 0.554 & \textbf{0.455} & 0.488 & 0.536 & 0.507 & 0.506 & 0.504\\ \hline
    \end{tabular}
    \label{crps}
\end{table}

\begin{table}[t!p]
    \centering
    \caption{Scaled Continuous Ranked Probability Score (sCRPS), ensuring that score comparisons remain meaningful across forecasts with different inherent variability. The stationary model achieves the lowest sCRPS for long maturities and long forecast horizons, while the spatiotemporal model with Gamma and lognormal priors performs best for short maturities and short horizons. For each row, the best model is highlighted in bold.}
    \begin{tabular}{l l l l l l l l l }
    \hline
        Horizons & Maturity & AR(1) & Stat & Nonstat & Aniso & Spatemp & Lognormal & Gamma \\ \hline
        1M & 3 months & 0.397 & 0.398 & 0.424 & 0.401 & \textbf{0.312} & 0.314 &	0.314 \\ 
        ~ & 1 year & 0.405 & 0.415 & 0.443 & 0.406 & \textbf{0.349} & 0.350 & 0.350 \\ 
        ~ & 3 years & 0.456 & 0.449 & 0.467 & 0.450 & 0.414 & \textbf{0.414}	& 0.414 \\ 
        ~ & 5 years & 0.463 & 0.432 & 0.457 & 0.456 & 0.408 & \textbf{0.406} & 0.408 \\ 
        ~ & 10 years & 0.394 & 0.390 & 0.433 & 0.390 & 0.369 & \textbf{0.367} & 0.368 \\ \hline
        6M & 3 months & 0.840 & 0.838 & 0.856 & 0.842 & 0.791 & 0.790 & \textbf{0.787} \\ 
        ~ & 1 year & 0.872 & 0.872 & 0.891 & 0.867 & 0.844 & 0.844 & \textbf{0.843} \\ 
        ~ & 3 years & 0.919 & \textbf{0.901} & 0.934 & 0.907 & 0.903 & 0.904	& 0.901 \\ 
        ~ & 5 years & 0.943 & \textbf{0.898} & 0.941 & 0.928 & 0.912 & 0.915 & 0.910 \\ 
        ~ & 10 years & 0.902 & \textbf{0.855} & 0.911 & 0.887 & 0.880 & 0.883 & 0.880 \\ \hline
        12M & 3 months & 1.022 & 1.013 & 0.994 & 1.022 & 0.991 & 0.987 & \textbf{0.983} \\ 
        ~ & 1 year & 1.023 & 1.009 & \textbf{0.993} & 1.020 & 1.008 & 1.006 & 1.002 \\
        ~ & 3 years & 1.038 & 0.996 & \textbf{0.993} & 1.028 & 1.028 & 1.026 & 1.024 \\ 
        ~ & 5 years & 1.060 & \textbf{0.996} & 1.007 & 1.047 & 1.039 & 1.039 & 1.036 \\ 
        ~ & 10 years & 1.054 & \textbf{0.966} & 1.001 & 1.038 & 1.019 & 1.019 & 1.017 \\ \hline
    \end{tabular}
    \label{scrps}
\end{table}

\begin{table}[t!p]
    \centering
    \caption{Weighted Continuous Ranked Probability Score (wCRPS) values, placing greater emphasis on forecast accuracy in the tails of the predictive distribution. The stationary model achieves the lowest wCRPS for long maturities and long forecast horizons, while the spatiotemporal model with Gamma and lognormal priors performs best for short maturities and short horizons. For each row, the best model is highlighted in bold.}
    \begin{tabular}{l l l l l l l l l }
    \hline
        Horizons & Maturity & AR(1) & Stat & Nonstat & Aniso & Spatemp & Lognormal & Gamma \\ \hline
        1M & 3 months & 0.015 & 0.012 & 0.023 & 0.016 & 0.007 & 0.007 & \textbf{0.007} \\ 
        ~ & 1 year & 0.011 & 0.012 & 0.024 & 0.011 & 0.008 & \textbf{0.008} & 0.008 \\ 
        ~ & 3 years & 0.020 & 0.018 & 0.026 & 0.020 & 0.016 & 0.016 & \textbf{0.016} \\ 
        ~ & 5 years & 0.020 & 0.017 & 0.025 & 0.020 & 0.016 & \textbf{0.016} & 0.016 \\ 
        ~ & 10 years & 0.014 & 0.012 & 0.023 & 0.014 & 0.012 & 0.012 & \textbf{0.012} \\ \hline
        6M & 3 months & 0.101 & 0.081 & 0.111 & 0.102 & 0.070 & 0.065 & \textbf{0.064} \\ 
        ~ & 1 year & 0.108 & 0.095 & 0.119 & 0.103 & 0.092 & 0.088 & \textbf{0.087} \\ 
        ~ & 3 years & 0.159 & \textbf{0.117} & 0.150 & 0.147 & 0.125 & 0.123 & 0.121 \\ 
        ~ & 5 years & 0.182 & \textbf{0.120} & 0.158 & 0.168 & 0.128 & 0.127	& 0.125 \\ 
        ~ & 10 years & 0.165 & \textbf{0.102} & 0.145 & 0.152 & 0.119 & 0.120 & 0.118 \\ \hline
        12M & 3 months & 0.166 & 0.123 & 0.154 & 0.165 & 0.114 & 0.108 & \textbf{0.106} \\ 
        ~ & 1 year & 0.168 & \textbf{0.123} & 0.151 & 0.159 & 0.131 & 0.126 & 0.125 \\ 
        ~ & 3 years & 0.226 & \textbf{0.143} & 0.177 & 0.209 & 0.167 & 0.164	& 0.163 \\ 
        ~ & 5 years & 0.268 & \textbf{0.153} & 0.198 & 0.248 & 0.181 & 0.180 & 0.179 \\ 
        ~ & 10 years & 0.292 & \textbf{0.150} & 0.207 & 0.271 & 0.191 & 0.193 & 0.192 \\ \hline
    \end{tabular}
    \label{wcrps}
\end{table}

\begin{table}[t!p]
    \centering
    \caption{Scaled and weighted Continuous Ranked Probability Score (swCRPS), a locally tail-scale invariant score to ensure fair evaluation of forecasts in these targeted regions, regardless of their scale. The stationary model achieves the lowest wCRPS for long maturities and long forecast horizons, while the spatiotemporal model with Gamma and lognormal priors performs best for short maturities and short horizons. For each row, the best model is highlighted in bold.} 
    \begin{tabular}{l l l l l l l l l }
    \hline
        Horizons & Maturity & AR(1) & Stat & Nonstat & Aniso & Spatemp & Lognormal & Gamma \\ \hline
        1M & 3 months & -0.386 & -0.439 & -0.382 & -0.378 & -0.656 & \textbf{-0.660} & -0.656 \\ 
        ~ & 1 year & -0.540 & -0.500 & -0.450 & -0.538 & -0.687 & \textbf{-0.699} & -0.694 \\ 
        ~ & 3 years & -0.427 & -0.514 & -0.471 & -0.440 & -0.627 & \textbf{-0.640} & -0.639 \\ 
        ~ & 5 years & -0.451 & -0.595 & -0.539 & -0.458 & -0.697 & -0.706 & \textbf{-0.710} \\ 
        ~ & 10 years & -0.638 & -0.735 & -0.648 & -0.645 & -0.831 & -0.833 & \textbf{-0.838} \\ \hline
        6M & 3 months & 0.284 & 0.191 & 0.271 & 0.289 & 0.130 & \textbf{0.103} & 0.113 \\ 
        ~ & 1 year & 0.193 & 0.111 & 0.175 & 0.184 & 0.083 & \textbf{0.052} & 0.065 \\ 
        ~ & 3 years & 0.215 & \textbf{0.061} & 0.138 & 0.191 & 0.071 & 0.043 & 0.054 \\ 
        ~ & 5 years & 0.226 & \textbf{0.020} & 0.103 & 0.197 & 0.046 & 0.021	& 0.031 \\ 
        ~ & 10 years & 0.213 & \textbf{-0.035} & 0.071 & 0.185 & 0.044 & 0.029 & 0.040 \\ \hline
        12M & 3 months & 0.448 & 0.300 & 0.352 & 0.452 & 0.293 & \textbf{0.262} & 0.271 \\ 
        ~ & 1 year & 0.375 & \textbf{0.208} & 0.251 & 0.368 & 0.246 & 0.219 & 0.225 \\ 
        ~ & 3 years & 0.435 & \textbf{0.195} & 0.260 & 0.409 & 0.276 & 0.252 & 0.264 \\ 
        ~ & 5 years & 0.465 & \textbf{0.168} & 0.247 & 0.438 & 0.265 & 0.250 & 0.256 \\ 
        ~ & 10 years & 0.527 & \textbf{0.175} & 0.269 & 0.496 & 0.314 & 0.315 & 0.317 \\ \hline
    \end{tabular}
    \label{swcrps}
\end{table}

The spatiotemporal model consistently outperforms others at short horizons and short-to-medium maturities, achieving the lowest CRPS values at 1-month ahead predictions. After scaling, Spatemp remains competitive, but differences between models amplify at longer horizons. The spatiotemporal model also excels in wCRPS and swCRPS, validating its ability to handle tail dependencies.

However, the stationary model demonstrates unexpected robustness for long-term maturities, likely due to its parsimony and stability. Conversely, the non-stationary model struggles universally, particularly in extreme regions and long horizons, cautioning against over-parameterization.

\subsection{Residual Correlation Diagnostics} \label{residualany}
To supplement the residual correlation analysis presented in the main text, we provide additional diagnostic evidence demonstrating the necessity of modeling residual dependence structure. We first examine the empirical variogram to visualize how residual correlation decays with distance in the time-maturity domain, and then apply formal statistical tests to quantify the strength of remaining dependence. Ideally, well-specified models should produce residuals with zero correlation, indicating that all systematic dependence has been captured by the model structure. 

\begin{figure}[t!]
    \centering
    \includegraphics[width=0.8\linewidth]{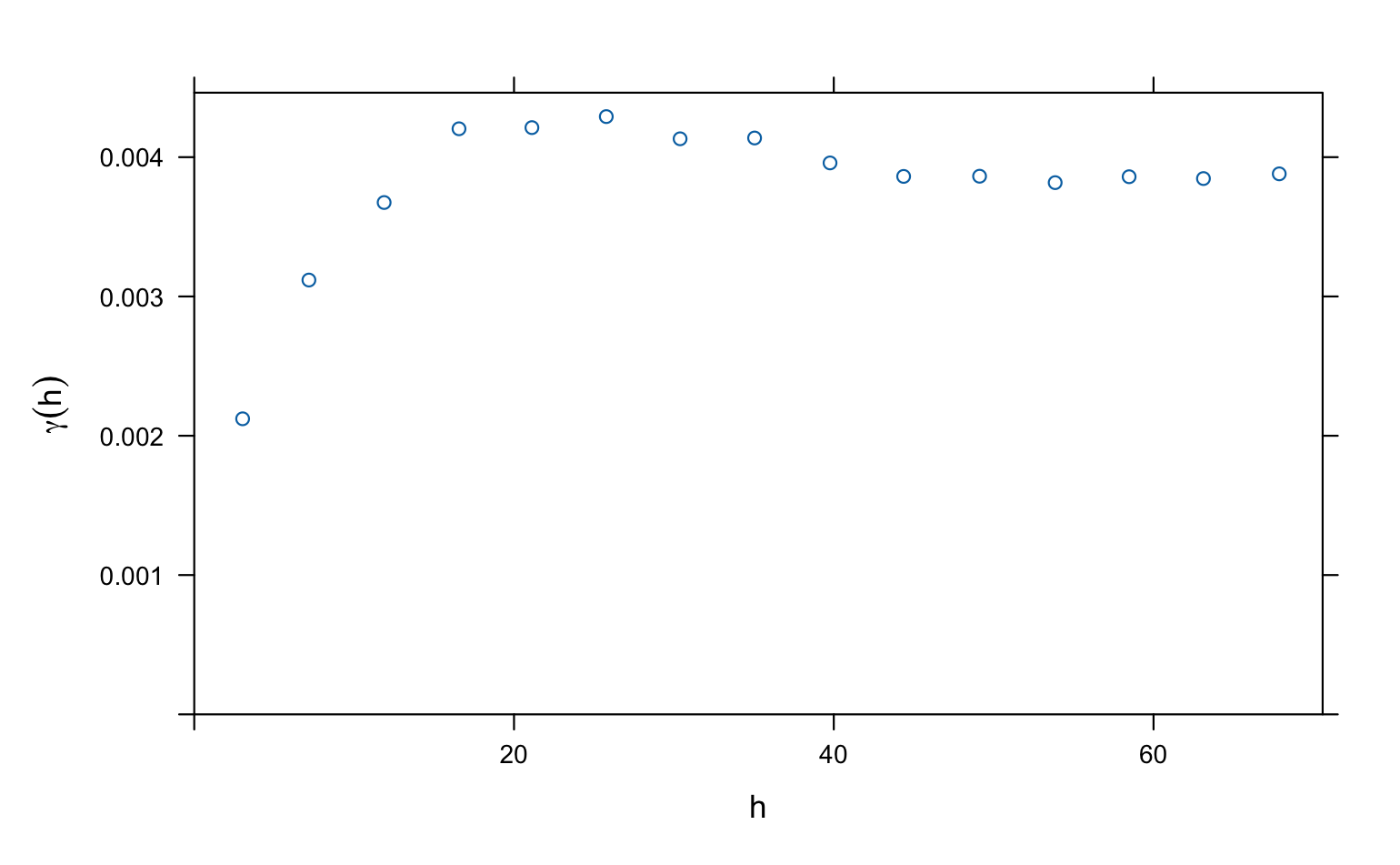}
    \caption{Sample variogram $\gamma(h)$ of the residuals of BDNS model, $h$ is the Euclidean distance between points in this 2D space, where time is treated as an ordered sequence of months (1 to 192) and maturity is on a log scale (log(3) to log(120)).}
    \label{fig:vario}
\end{figure}

The empirical variogram of the residuals of BDNS model (Figure~\ref{fig:vario}) offers a continuous view of "spatial" dependence along the maturity axis. After log-transforming time to balance spacing (\citeay{zhuang2021statistical}), the variogram increases smoothly with "distance" in the (time, log-maturity) domain before reaching a clear sill, indicating that residual correlation decays gradually rather than disappearing abruptly. This behavior is consistent with the clustered patterns observed in the correlation heatmaps and with a latent Gaussian random field exhibiting finite smoothness. Together, these features indicate that the residual process retains structured spatial dependence.

BDNS residual correlations across time display a pronounced autoregressive pattern: high short-lag correlations that decay only gradually. This reflects persistence in short-term rate shocks and slow mean reversion in long-term yields. The average lag-1 autocorrelation across maturities is 0.585. Hence, although the latent factors are modeled as AR(1) processes, their innovations remain serially correlated, implying that a single common AR component is insufficient to capture the temporal dependence of yield-curve residuals.

``Spatial'' dependence measures further support this conclusion. The average Moran’s \( I \) statistic across maturities equals 0.284 (SD = 0.275), with approximately 47\% of time periods showing significant spatial correlation (\( \text{p-value} < 0.05 \)). This moderate but persistent positive autocorrelation indicates clustering of large residuals among neighboring maturities. The average Geary’s \( C \) statistic equals 0.655, a value consistent with positive local dependence. Together, these diagnostics confirm that the residual field exhibits ``spatial'' smoothness inconsistent with the independence assumption, and that cross-maturity dependence remains a key source of model misspecification.

Overall, the BDNS model produces residuals that are neither temporally nor spatially uncorrelated. The persistence and cross-maturity clustering suggest that the latent BDNS structure is too rigid to capture the bivariate dynamics of yield-curve innovations. These observations motivate modeling residual dependence explicitly across time and maturity, which we address using SPDE-based extensions that induce flexible Matérn-type dependence structures. 

The spatiotemporal model shows the most progress toward residual decorrelation among the SPDE specifications examined, though it remains imperfect. Across maturities, while red regions persist, the overall intensity of off-diagonal correlations appears reduced compared to other SPDE models. Across time, the temporal correlation structure shows the most improvement relative to the alternatives, with a cleaner diagonal pattern and reduced autocorrelation intensity. While the joint modeling of spatial and temporal dynamics does not achieve the ideal white noise structure, it represents the best compromise among the available SPDE formulations for capturing the bivariate dependence in this application.

Table~\ref{residual} summarizes the residual dependence diagnostics across maturities and time for the baseline BDNS model (AR(1)) and the three SPDE-based extensions. The AR(1) specification leaves pronounced spatial and temporal dependence, as indicated by high average absolute correlations (0.29), positive Moran’s \( I = 0.29 \), and a mean lag-1 autocorrelation of 0.58. In contrast, all SPDE formulations substantially reduce both forms of dependence, yet they differ slightly in strength and balance. The \textit{Spatemp} model, which incorporates both spatial and temporal random-field effects, yields the largest reduction in overall correlation (Abs.\ Corr.\ = 0.1184) and produces mildly negative spatial dependence (Moran’s \( I = -0.27 \)) together with an increase in Geary’s \( C = 1.19 \), consistent with locally alternating residuals. The \textit{Gamma} prior version achieves nearly identical spatial metrics but exhibits slightly less temporal whitening (ACF(1) = -0.19), suggesting that its heavier-tailed prior induces marginally weaker shrinkage in time. The \textit{Lognormal} variant performs almost identically to \textit{Spatemp}, with marginally higher absolute correlation and nearly the same spatial statistics, confirming that the results are robust to the log-scale parameterization of the range prior. Overall, the SPDE-based residual specifications effectively eliminate the strong cross-maturity and serial dependence present in the BDNS model, while small differences across priors indicate that model performance is stable with respect to the choice of hyperprior on the spatial precision parameter.

\begin{table}[t!]
    \centering
    \caption{Residual dependence statistics for the BDNS model and SPDE-based extensions. Lower absolute correlation, negative Moran's I values, elevated Geary's C, and negative first-order autocorrelation (ACF1) indicate the SPDE models effectively captured the covariance structure of residuals}
    \begin{tabular}{|c|c|c|c|c|}
    \hline
        Model &  Abs Corr & Moran's I & Geary’s C & ACF1\\ \hline
        AR(1) &   0.2866 & 0.2853 & 0.6551 & 0.5843 \\ \hline
        Spatemp & 0.1184 & -0.2703 & 1.1918 & -0.2026 \\ \hline
        Gamma & 0.1175 & -0.2683 & 1.1900 & -0.1905 \\ \hline
        Lognormal & 0.1188 & -0.2715 & 1.1933 & -0.2012 \\ \hline
    \end{tabular}
    \label{residual}
\end{table}

\end{document}